\documentclass[11pt,a4paper,english,superscriptaddress,aps,preprint]{revtex4}

\usepackage{color}

\usepackage{amsmath}
\usepackage{amssymb}
\usepackage{graphicx}
\usepackage{hhline}
\usepackage{textcomp}
\makeatletter
\usepackage{babel}
\newcommand{\bea}{\begin{eqnarray}}
\newcommand{\eea}{\end{eqnarray}}

\newcommand{\pa}{\partial}
\renewcommand{\a}{\alpha}

\renewcommand{\b}{\beta}

\newcommand{\q}{\theta}

\newcommand{\be}{\begin{equation}}
\newcommand{\ee}{\end{equation}}
\newtheorem{theorem}{Theorem}
\newtheorem{lemma}[theorem]{Lemma}

\newcommand{\qed}{\nobreak \ifvmode \relax \else
      \ifdim\lastskip<1.5em \hskip-\lastskip
      \hskip1.5em plus0em minus0.5em \fi \nobreak
      \vrule height0.75em width0.5em depth0.25em\fi}

\begin{document}
\immediate\write16{<<WARNING: LINEDRAW macros work with emTeX-dvivers
                    and other drivers supporting emTeX \special's
                    (dviscr, dvihplj, dvidot, dvips, dviwin, etc.) >>}

\newdimen\Lengthunit       \Lengthunit  = 1.5cm
\newcount\Nhalfperiods     \Nhalfperiods= 9
\newcount\magnitude        \magnitude = 1000

\catcode`\*=11
\newdimen\L*   \newdimen\d*   \newdimen\d**
\newdimen\dm*  \newdimen\dd*  \newdimen\dt*
\newdimen\a*   \newdimen\b*   \newdimen\c*
\newdimen\a**  \newdimen\b**
\newdimen\xL*  \newdimen\yL*
\newdimen\rx*  \newdimen\ry*
\newdimen\tmp* \newdimen\linwid*

\newcount\k*   \newcount\l*   \newcount\m*
\newcount\k**  \newcount\l**  \newcount\m**
\newcount\n*   \newcount\dn*  \newcount\r*
\newcount\N*   \newcount\*one \newcount\*two  \*one=1 \*two=2
\newcount\*ths \*ths=1000
\newcount\angle*  \newcount\q*  \newcount\q**
\newcount\angle** \angle**=0
\newcount\sc*     \sc*=0

\newtoks\cos*  \cos*={1}
\newtoks\sin*  \sin*={0}

\catcode`\[=13

\def\rotate(#1){\advance\angle**#1\angle*=\angle**
\q**=\angle*\ifnum\q**<0\q**=-\q**\fi
\ifnum\q**>360\q*=\angle*\divide\q*360\multiply\q*360\advance\angle*-\q*\fi
\ifnum\angle*<0\advance\angle*360\fi\q**=\angle*\divide\q**90\q**=\q**
\def\sgcos*{+}\def\sgsin*{+}\relax
\ifcase\q**\or
 \def\sgcos*{-}\def\sgsin*{+}\or
 \def\sgcos*{-}\def\sgsin*{-}\or
 \def\sgcos*{+}\def\sgsin*{-}\else\fi
\q*=\q**
\multiply\q*90\advance\angle*-\q*
\ifnum\angle*>45\sc*=1\angle*=-\angle*\advance\angle*90\else\sc*=0\fi
\def[##1,##2]{\ifnum\sc*=0\relax
\edef\cs*{\sgcos*.##1}\edef\sn*{\sgsin*.##2}\ifcase\q**\or
 \edef\cs*{\sgcos*.##2}\edef\sn*{\sgsin*.##1}\or
 \edef\cs*{\sgcos*.##1}\edef\sn*{\sgsin*.##2}\or
 \edef\cs*{\sgcos*.##2}\edef\sn*{\sgsin*.##1}\else\fi\else
\edef\cs*{\sgcos*.##2}\edef\sn*{\sgsin*.##1}\ifcase\q**\or
 \edef\cs*{\sgcos*.##1}\edef\sn*{\sgsin*.##2}\or
 \edef\cs*{\sgcos*.##2}\edef\sn*{\sgsin*.##1}\or
 \edef\cs*{\sgcos*.##1}\edef\sn*{\sgsin*.##2}\else\fi\fi
\cos*={\cs*}\sin*={\sn*}\global\edef\gcos*{\cs*}\global\edef\gsin*{\sn*}}\relax
\ifcase\angle*[9999,0]\or
[999,017]\or[999,034]\or[998,052]\or[997,069]\or[996,087]\or
[994,104]\or[992,121]\or[990,139]\or[987,156]\or[984,173]\or
[981,190]\or[978,207]\or[974,224]\or[970,241]\or[965,258]\or
[961,275]\or[956,292]\or[951,309]\or[945,325]\or[939,342]\or
[933,358]\or[927,374]\or[920,390]\or[913,406]\or[906,422]\or
[898,438]\or[891,453]\or[882,469]\or[874,484]\or[866,499]\or
[857,515]\or[848,529]\or[838,544]\or[829,559]\or[819,573]\or
[809,587]\or[798,601]\or[788,615]\or[777,629]\or[766,642]\or
[754,656]\or[743,669]\or[731,681]\or[719,694]\or[707,707]\or
\else[9999,0]\fi}

\catcode`\[=12

\def\GRAPH(hsize=#1)#2{\hbox to #1\Lengthunit{#2\hss}}

\def\Linewidth#1{\global\linwid*=#1\relax
\global\divide\linwid*10\global\multiply\linwid*\mag
\global\divide\linwid*100\special{em:linewidth \the\linwid*}}

\Linewidth{.4pt}
\def\sm*{\special{em:moveto}}
\def\sl*{\special{em:lineto}}
\let\moveto=\sm*
\let\lineto=\sl*
\newbox\spm*   \newbox\spl*
\setbox\spm*\hbox{\sm*}
\setbox\spl*\hbox{\sl*}

\def\mov#1(#2,#3)#4{\rlap{\L*=#1\Lengthunit
\xL*=#2\L* \yL*=#3\L*
\xL*=\xscale\xL* \yL*=\yscale\yL*
\rx* \the\cos*\xL* \tmp* \the\sin*\yL* \advance\rx*-\tmp*
\ry* \the\cos*\yL* \tmp* \the\sin*\xL* \advance\ry*\tmp*
\kern\rx*\raise\ry*\hbox{#4}}}

\def\rmov*(#1,#2)#3{\rlap{\xL*=#1\yL*=#2\relax
\rx* \the\cos*\xL* \tmp* \the\sin*\yL* \advance\rx*-\tmp*
\ry* \the\cos*\yL* \tmp* \the\sin*\xL* \advance\ry*\tmp*
\kern\rx*\raise\ry*\hbox{#3}}}

\def\lin#1(#2,#3){\rlap{\sm*\mov#1(#2,#3){\sl*}}}

\def\arr*(#1,#2,#3){\rmov*(#1\dd*,#1\dt*){\sm*
\rmov*(#2\dd*,#2\dt*){\rmov*(#3\dt*,-#3\dd*){\sl*}}\sm*
\rmov*(#2\dd*,#2\dt*){\rmov*(-#3\dt*,#3\dd*){\sl*}}}}

\def\arrow#1(#2,#3){\rlap{\lin#1(#2,#3)\mov#1(#2,#3){\relax
\d**=-.012\Lengthunit\dd*=#2\d**\dt*=#3\d**
\arr*(1,10,4)\arr*(3,8,4)\arr*(4.8,4.2,3)}}}

\def\arrlin#1(#2,#3){\rlap{\L*=#1\Lengthunit\L*=.5\L*
\lin#1(#2,#3)\rmov*(#2\L*,#3\L*){\arrow.1(#2,#3)}}}

\def\dasharrow#1(#2,#3){\rlap{{\Lengthunit=0.9\Lengthunit
\dashlin#1(#2,#3)\mov#1(#2,#3){\sm*}}\mov#1(#2,#3){\sl*
\d**=-.012\Lengthunit\dd*=#2\d**\dt*=#3\d**
\arr*(1,10,4)\arr*(3,8,4)\arr*(4.8,4.2,3)}}}

\def\clap#1{\hbox to 0pt{\hss #1\hss}}

\def\ind(#1,#2)#3{\rlap{\L*=.1\Lengthunit
\xL*=#1\L* \yL*=#2\L*
\rx* \the\cos*\xL* \tmp* \the\sin*\yL* \advance\rx*-\tmp*
\ry* \the\cos*\yL* \tmp* \the\sin*\xL* \advance\ry*\tmp*
\kern\rx*\raise\ry*\hbox{\lower2pt\clap{$#3$}}}}

\def\sh*(#1,#2)#3{\rlap{\dm*=\the\n*\d**
\xL*=\xscale\dm* \yL*=\yscale\dm* \xL*=#1\xL* \yL*=#2\yL*
\rx* \the\cos*\xL* \tmp* \the\sin*\yL* \advance\rx*-\tmp*
\ry* \the\cos*\yL* \tmp* \the\sin*\xL* \advance\ry*\tmp*
\kern\rx*\raise\ry*\hbox{#3}}}

\def\calcnum*#1(#2,#3){\a*=1000sp\b*=1000sp\a*=#2\a*\b*=#3\b*
\ifdim\a*<0pt\a*-\a*\fi\ifdim\b*<0pt\b*-\b*\fi
\ifdim\a*>\b*\c*=.96\a*\advance\c*.4\b*
\else\c*=.96\b*\advance\c*.4\a*\fi
\k*\a*\multiply\k*\k*\l*\b*\multiply\l*\l*
\m*\k*\advance\m*\l*\n*\c*\r*\n*\multiply\n*\n*
\dn*\m*\advance\dn*-\n*\divide\dn*2\divide\dn*\r*
\advance\r*\dn*
\c*=\the\Nhalfperiods5sp\c*=#1\c*\ifdim\c*<0pt\c*-\c*\fi
\multiply\c*\r*\N*\c*\divide\N*10000}

\def\dashlin#1(#2,#3){\rlap{\calcnum*#1(#2,#3)\relax
\d**=#1\Lengthunit\ifdim\d**<0pt\d**-\d**\fi
\divide\N*2\multiply\N*2\advance\N*\*one
\divide\d**\N*\sm*\n*\*one\sh*(#2,#3){\sl*}\loop
\advance\n*\*one\sh*(#2,#3){\sm*}\advance\n*\*one
\sh*(#2,#3){\sl*}\ifnum\n*<\N*\repeat}}

\def\dashdotlin#1(#2,#3){\rlap{\calcnum*#1(#2,#3)\relax
\d**=#1\Lengthunit\ifdim\d**<0pt\d**-\d**\fi
\divide\N*2\multiply\N*2\advance\N*1\multiply\N*2\relax
\divide\d**\N*\sm*\n*\*two\sh*(#2,#3){\sl*}\loop
\advance\n*\*one\sh*(#2,#3){\kern-1.48pt\lower.5pt\hbox{\rm.}}\relax
\advance\n*\*one\sh*(#2,#3){\sm*}\advance\n*\*two
\sh*(#2,#3){\sl*}\ifnum\n*<\N*\repeat}}

\def\shl*(#1,#2)#3{\kern#1#3\lower#2#3\hbox{\unhcopy\spl*}}

\def\trianglin#1(#2,#3){\rlap{\toks0={#2}\toks1={#3}\calcnum*#1(#2,#3)\relax
\dd*=.57\Lengthunit\dd*=#1\dd*\divide\dd*\N*
\divide\dd*\*ths \multiply\dd*\magnitude
\d**=#1\Lengthunit\ifdim\d**<0pt\d**-\d**\fi
\multiply\N*2\divide\d**\N*\sm*\n*\*one\loop
\shl**{\dd*}\dd*-\dd*\advance\n*2\relax
\ifnum\n*<\N*\repeat\n*\N*\shl**{0pt}}}

\def\wavelin#1(#2,#3){\rlap{\toks0={#2}\toks1={#3}\calcnum*#1(#2,#3)\relax
\dd*=.23\Lengthunit\dd*=#1\dd*\divide\dd*\N*
\divide\dd*\*ths \multiply\dd*\magnitude
\d**=#1\Lengthunit\ifdim\d**<0pt\d**-\d**\fi
\multiply\N*4\divide\d**\N*\sm*\n*\*one\loop
\shl**{\dd*}\dt*=1.3\dd*\advance\n*\*one
\shl**{\dt*}\advance\n*\*one
\shl**{\dd*}\advance\n*\*two
\dd*-\dd*\ifnum\n*<\N*\repeat\n*\N*\shl**{0pt}}}

\def\w*lin(#1,#2){\rlap{\toks0={#1}\toks1={#2}\d**=\Lengthunit\dd*=-.12\d**
\divide\dd*\*ths \multiply\dd*\magnitude
\N*8\divide\d**\N*\sm*\n*\*one\loop
\shl**{\dd*}\dt*=1.3\dd*\advance\n*\*one
\shl**{\dt*}\advance\n*\*one
\shl**{\dd*}\advance\n*\*one
\shl**{0pt}\dd*-\dd*\advance\n*1\ifnum\n*<\N*\repeat}}

\def\l*arc(#1,#2)[#3][#4]{\rlap{\toks0={#1}\toks1={#2}\d**=\Lengthunit
\dd*=#3.037\d**\dd*=#4\dd*\dt*=#3.049\d**\dt*=#4\dt*\ifdim\d**>10mm\relax
\d**=.25\d**\n*\*one\shl**{-\dd*}\n*\*two\shl**{-\dt*}\n*3\relax
\shl**{-\dd*}\n*4\relax\shl**{0pt}\else
\ifdim\d**>5mm\d**=.5\d**\n*\*one\shl**{-\dt*}\n*\*two
\shl**{0pt}\else\n*\*one\shl**{0pt}\fi\fi}}

\def\d*arc(#1,#2)[#3][#4]{\rlap{\toks0={#1}\toks1={#2}\d**=\Lengthunit
\dd*=#3.037\d**\dd*=#4\dd*\d**=.25\d**\sm*\n*\*one\shl**{-\dd*}\relax
\n*3\relax\sh*(#1,#2){\xL*=\xscale\dd*\yL*=\yscale\dd*
\kern#2\xL*\lower#1\yL*\hbox{\sm*}}\n*4\relax\shl**{0pt}}}

\def\shl**#1{\c*=\the\n*\d**\d*=#1\relax
\a*=\the\toks0\c*\b*=\the\toks1\d*\advance\a*-\b*
\b*=\the\toks1\c*\d*=\the\toks0\d*\advance\b*\d*
\a*=\xscale\a*\b*=\yscale\b*
\rx* \the\cos*\a* \tmp* \the\sin*\b* \advance\rx*-\tmp*
\ry* \the\cos*\b* \tmp* \the\sin*\a* \advance\ry*\tmp*
\raise\ry*\rlap{\kern\rx*\unhcopy\spl*}}

\def\wlin*#1(#2,#3)[#4]{\rlap{\toks0={#2}\toks1={#3}\relax
\c*=#1\l*\c*\c*=.01\Lengthunit\m*\c*\divide\l*\m*
\c*=\the\Nhalfperiods5sp\multiply\c*\l*\N*\c*\divide\N*\*ths
\divide\N*2\multiply\N*2\advance\N*\*one
\dd*=.002\Lengthunit\dd*=#4\dd*\multiply\dd*\l*\divide\dd*\N*
\divide\dd*\*ths \multiply\dd*\magnitude
\d**=#1\multiply\N*4\divide\d**\N*\sm*\n*\*one\loop
\shl**{\dd*}\dt*=1.3\dd*\advance\n*\*one
\shl**{\dt*}\advance\n*\*one
\shl**{\dd*}\advance\n*\*two
\dd*-\dd*\ifnum\n*<\N*\repeat\n*\N*\shl**{0pt}}}

\def\wavebox#1{\setbox0\hbox{#1}\relax
\a*=\wd0\advance\a*14pt\b*=\ht0\advance\b*\dp0\advance\b*14pt\relax
\hbox{\kern9pt\relax
\rmov*(0pt,\ht0){\rmov*(-7pt,7pt){\wlin*\a*(1,0)[+]\wlin*\b*(0,-1)[-]}}\relax
\rmov*(\wd0,-\dp0){\rmov*(7pt,-7pt){\wlin*\a*(-1,0)[+]\wlin*\b*(0,1)[-]}}\relax
\box0\kern9pt}}

\def\rectangle#1(#2,#3){\relax
\lin#1(#2,0)\lin#1(0,#3)\mov#1(0,#3){\lin#1(#2,0)}\mov#1(#2,0){\lin#1(0,#3)}}

\def\dashrectangle#1(#2,#3){\dashlin#1(#2,0)\dashlin#1(0,#3)\relax
\mov#1(0,#3){\dashlin#1(#2,0)}\mov#1(#2,0){\dashlin#1(0,#3)}}

\def\waverectangle#1(#2,#3){\L*=#1\Lengthunit\a*=#2\L*\b*=#3\L*
\ifdim\a*<0pt\a*-\a*\def\x*{-1}\else\def\x*{1}\fi
\ifdim\b*<0pt\b*-\b*\def\y*{-1}\else\def\y*{1}\fi
\wlin*\a*(\x*,0)[-]\wlin*\b*(0,\y*)[+]\relax
\mov#1(0,#3){\wlin*\a*(\x*,0)[+]}\mov#1(#2,0){\wlin*\b*(0,\y*)[-]}}

\def\calcparab*{\ifnum\n*>\m*\k*\N*\advance\k*-\n*\else\k*\n*\fi
\a*=\the\k* sp\a*=10\a*\b*\dm*\advance\b*-\a*\k*\b*
\a*=\the\*ths\b*\divide\a*\l*\multiply\a*\k*
\divide\a*\l*\k*\*ths\r*\a*\advance\k*-\r*\dt*=\the\k*\L*}

\def\arcto#1(#2,#3)[#4]{\rlap{\toks0={#2}\toks1={#3}\calcnum*#1(#2,#3)\relax
\dm*=135sp\dm*=#1\dm*\d**=#1\Lengthunit\ifdim\dm*<0pt\dm*-\dm*\fi
\multiply\dm*\r*\a*=.3\dm*\a*=#4\a*\ifdim\a*<0pt\a*-\a*\fi
\advance\dm*\a*\N*\dm*\divide\N*10000\relax
\divide\N*2\multiply\N*2\advance\N*\*one
\L*=-.25\d**\L*=#4\L*\divide\d**\N*\divide\L*\*ths
\m*\N*\divide\m*2\dm*=\the\m*5sp\l*\dm*\sm*\n*\*one\loop
\calcparab*\shl**{-\dt*}\advance\n*1\ifnum\n*<\N*\repeat}}

\def\arrarcto#1(#2,#3)[#4]{\L*=#1\Lengthunit\L*=.54\L*
\arcto#1(#2,#3)[#4]\rmov*(#2\L*,#3\L*){\d*=.457\L*\d*=#4\d*\d**-\d*
\rmov*(#3\d**,#2\d*){\arrow.02(#2,#3)}}}

\def\dasharcto#1(#2,#3)[#4]{\rlap{\toks0={#2}\toks1={#3}\relax
\calcnum*#1(#2,#3)\dm*=\the\N*5sp\a*=.3\dm*\a*=#4\a*\ifdim\a*<0pt\a*-\a*\fi
\advance\dm*\a*\N*\dm*
\divide\N*20\multiply\N*2\advance\N*1\d**=#1\Lengthunit
\L*=-.25\d**\L*=#4\L*\divide\d**\N*\divide\L*\*ths
\m*\N*\divide\m*2\dm*=\the\m*5sp\l*\dm*
\sm*\n*\*one\loop\calcparab*
\shl**{-\dt*}\advance\n*1\ifnum\n*>\N*\else\calcparab*
\sh*(#2,#3){\xL*=#3\dt* \yL*=#2\dt*
\rx* \the\cos*\xL* \tmp* \the\sin*\yL* \advance\rx*\tmp*
\ry* \the\cos*\yL* \tmp* \the\sin*\xL* \advance\ry*-\tmp*
\kern\rx*\lower\ry*\hbox{\sm*}}\fi
\advance\n*1\ifnum\n*<\N*\repeat}}

\def\*shl*#1{\c*=\the\n*\d**\advance\c*#1\a**\d*\dt*\advance\d*#1\b**
\a*=\the\toks0\c*\b*=\the\toks1\d*\advance\a*-\b*
\b*=\the\toks1\c*\d*=\the\toks0\d*\advance\b*\d*
\rx* \the\cos*\a* \tmp* \the\sin*\b* \advance\rx*-\tmp*
\ry* \the\cos*\b* \tmp* \the\sin*\a* \advance\ry*\tmp*
\raise\ry*\rlap{\kern\rx*\unhcopy\spl*}}

\def\calcnormal*#1{\b**=10000sp\a**\b**\k*\n*\advance\k*-\m*
\multiply\a**\k*\divide\a**\m*\a**=#1\a**\ifdim\a**<0pt\a**-\a**\fi
\ifdim\a**>\b**\d*=.96\a**\advance\d*.4\b**
\else\d*=.96\b**\advance\d*.4\a**\fi
\d*=.01\d*\r*\d*\divide\a**\r*\divide\b**\r*
\ifnum\k*<0\a**-\a**\fi\d*=#1\d*\ifdim\d*<0pt\b**-\b**\fi
\k*\a**\a**=\the\k*\dd*\k*\b**\b**=\the\k*\dd*}

\def\wavearcto#1(#2,#3)[#4]{\rlap{\toks0={#2}\toks1={#3}\relax
\calcnum*#1(#2,#3)\c*=\the\N*5sp\a*=.4\c*\a*=#4\a*\ifdim\a*<0pt\a*-\a*\fi
\advance\c*\a*\N*\c*\divide\N*20\multiply\N*2\advance\N*-1\multiply\N*4\relax
\d**=#1\Lengthunit\dd*=.012\d**
\divide\dd*\*ths \multiply\dd*\magnitude
\ifdim\d**<0pt\d**-\d**\fi\L*=.25\d**
\divide\d**\N*\divide\dd*\N*\L*=#4\L*\divide\L*\*ths
\m*\N*\divide\m*2\dm*=\the\m*0sp\l*\dm*
\sm*\n*\*one\loop\calcnormal*{#4}\calcparab*
\*shl*{1}\advance\n*\*one\calcparab*
\*shl*{1.3}\advance\n*\*one\calcparab*
\*shl*{1}\advance\n*2\dd*-\dd*\ifnum\n*<\N*\repeat\n*\N*\shl**{0pt}}}

\def\triangarcto#1(#2,#3)[#4]{\rlap{\toks0={#2}\toks1={#3}\relax
\calcnum*#1(#2,#3)\c*=\the\N*5sp\a*=.4\c*\a*=#4\a*\ifdim\a*<0pt\a*-\a*\fi
\advance\c*\a*\N*\c*\divide\N*20\multiply\N*2\advance\N*-1\multiply\N*2\relax
\d**=#1\Lengthunit\dd*=.012\d**
\divide\dd*\*ths \multiply\dd*\magnitude
\ifdim\d**<0pt\d**-\d**\fi\L*=.25\d**
\divide\d**\N*\divide\dd*\N*\L*=#4\L*\divide\L*\*ths
\m*\N*\divide\m*2\dm*=\the\m*0sp\l*\dm*
\sm*\n*\*one\loop\calcnormal*{#4}\calcparab*
\*shl*{1}\advance\n*2\dd*-\dd*\ifnum\n*<\N*\repeat\n*\N*\shl**{0pt}}}

\def\hr*#1{\L*=\xscale\Lengthunit\ifnum
\angle**=0\clap{\vrule width#1\L* height.1pt}\else
\L*=#1\L*\L*=.5\L*\rmov*(-\L*,0pt){\sm*}\rmov*(\L*,0pt){\sl*}\fi}

\def\shade#1[#2]{\rlap{\Lengthunit=#1\Lengthunit
\special{em:linewidth .001pt}\relax
\mov(0,#2.05){\hr*{.994}}\mov(0,#2.1){\hr*{.980}}\relax
\mov(0,#2.15){\hr*{.953}}\mov(0,#2.2){\hr*{.916}}\relax
\mov(0,#2.25){\hr*{.867}}\mov(0,#2.3){\hr*{.798}}\relax
\mov(0,#2.35){\hr*{.715}}\mov(0,#2.4){\hr*{.603}}\relax
\mov(0,#2.45){\hr*{.435}}\special{em:linewidth \the\linwid*}}}

\def\dshade#1[#2]{\rlap{\special{em:linewidth .001pt}\relax
\Lengthunit=#1\Lengthunit\if#2-\def\t*{+}\else\def\t*{-}\fi
\mov(0,\t*.025){\relax
\mov(0,#2.05){\hr*{.995}}\mov(0,#2.1){\hr*{.988}}\relax
\mov(0,#2.15){\hr*{.969}}\mov(0,#2.2){\hr*{.937}}\relax
\mov(0,#2.25){\hr*{.893}}\mov(0,#2.3){\hr*{.836}}\relax
\mov(0,#2.35){\hr*{.760}}\mov(0,#2.4){\hr*{.662}}\relax
\mov(0,#2.45){\hr*{.531}}\mov(0,#2.5){\hr*{.320}}\relax
\special{em:linewidth \the\linwid*}}}}

\def\vdot{\rlap{\kern-1.9pt\lower1.8pt\hbox{$\scriptstyle\bullet$}}}
\def\vtimes{\rlap{\kern-3pt\lower1.8pt\hbox{$\scriptstyle\times$}}}
\def\vDot{\rlap{\kern-2.3pt\lower2.7pt\hbox{$\bullet$}}}
\def\vTimes{\rlap{\kern-3.6pt\lower2.4pt\hbox{$\times$}}}

\def\arc(#1)[#2,#3]{{\k*=#2\l*=#3\m*=\l*
\advance\m*-6\ifnum\k*>\l*\relax\else
{\rotate(#2)\mov(#1,0){\sm*}}\loop
\ifnum\k*<\m*\advance\k*5{\rotate(\k*)\mov(#1,0){\sl*}}\repeat
{\rotate(#3)\mov(#1,0){\sl*}}\fi}}

\def\dasharc(#1)[#2,#3]{{\k**=#2\n*=#3\advance\n*-1\advance\n*-\k**
\L*=1000sp\L*#1\L* \multiply\L*\n* \multiply\L*\Nhalfperiods
\divide\L*57\N*\L* \divide\N*2000\ifnum\N*=0\N*1\fi
\r*\n*  \divide\r*\N* \ifnum\r*<2\r*2\fi
\m**\r* \divide\m**2 \l**\r* \advance\l**-\m** \N*\n* \divide\N*\r*
\k**\r* \multiply\k**\N* \dn*\n*
\advance\dn*-\k** \divide\dn*2\advance\dn*\*one
\r*\l** \divide\r*2\advance\dn*\r* \advance\N*-2\k**#2\relax
\ifnum\l**<6{\rotate(#2)\mov(#1,0){\sm*}}\advance\k**\dn*
{\rotate(\k**)\mov(#1,0){\sl*}}\advance\k**\m**
{\rotate(\k**)\mov(#1,0){\sm*}}\loop
\advance\k**\l**{\rotate(\k**)\mov(#1,0){\sl*}}\advance\k**\m**
{\rotate(\k**)\mov(#1,0){\sm*}}\advance\N*-1\ifnum\N*>0\repeat
{\rotate(#3)\mov(#1,0){\sl*}}\else\advance\k**\dn*
\arc(#1)[#2,\k**]\loop\advance\k**\m** \r*\k**
\advance\k**\l** {\arc(#1)[\r*,\k**]}\relax
\advance\N*-1\ifnum\N*>0\repeat
\advance\k**\m**\arc(#1)[\k**,#3]\fi}}

\def\triangarc#1(#2)[#3,#4]{{\k**=#3\n*=#4\advance\n*-\k**
\L*=1000sp\L*#2\L* \multiply\L*\n* \multiply\L*\Nhalfperiods
\divide\L*57\N*\L* \divide\N*1000\ifnum\N*=0\N*1\fi
\d**=#2\Lengthunit \d*\d** \divide\d*57\multiply\d*\n*
\r*\n*  \divide\r*\N* \ifnum\r*<2\r*2\fi
\m**\r* \divide\m**2 \l**\r* \advance\l**-\m** \N*\n* \divide\N*\r*
\dt*\d* \divide\dt*\N* \dt*.5\dt* \dt*#1\dt*
\divide\dt*1000\multiply\dt*\magnitude
\k**\r* \multiply\k**\N* \dn*\n* \advance\dn*-\k** \divide\dn*2\relax
\r*\l** \divide\r*2\advance\dn*\r* \advance\N*-1\k**#3\relax
{\rotate(#3)\mov(#2,0){\sm*}}\advance\k**\dn*
{\rotate(\k**)\mov(#2,0){\sl*}}\advance\k**-\m**\advance\l**\m**\loop\dt*-\dt*
\d*\d** \advance\d*\dt*
\advance\k**\l**{\rotate(\k**)\rmov*(\d*,0pt){\sl*}}%
\advance\N*-1\ifnum\N*>0\repeat\advance\k**\m**
{\rotate(\k**)\mov(#2,0){\sl*}}{\rotate(#4)\mov(#2,0){\sl*}}}}

\def\wavearc#1(#2)[#3,#4]{{\k**=#3\n*=#4\advance\n*-\k**
\L*=4000sp\L*#2\L* \multiply\L*\n* \multiply\L*\Nhalfperiods
\divide\L*57\N*\L* \divide\N*1000\ifnum\N*=0\N*1\fi
\d**=#2\Lengthunit \d*\d** \divide\d*57\multiply\d*\n*
\r*\n*  \divide\r*\N* \ifnum\r*=0\r*1\fi
\m**\r* \divide\m**2 \l**\r* \advance\l**-\m** \N*\n* \divide\N*\r*
\dt*\d* \divide\dt*\N* \dt*.7\dt* \dt*#1\dt*
\divide\dt*1000\multiply\dt*\magnitude
\k**\r* \multiply\k**\N* \dn*\n* \advance\dn*-\k** \divide\dn*2\relax
\divide\N*4\advance\N*-1\k**#3\relax
{\rotate(#3)\mov(#2,0){\sm*}}\advance\k**\dn*
{\rotate(\k**)\mov(#2,0){\sl*}}\advance\k**-\m**\advance\l**\m**\loop\dt*-\dt*
\d*\d** \advance\d*\dt* \dd*\d** \advance\dd*1.3\dt*
\advance\k**\r*{\rotate(\k**)\rmov*(\d*,0pt){\sl*}}\relax
\advance\k**\r*{\rotate(\k**)\rmov*(\dd*,0pt){\sl*}}\relax
\advance\k**\r*{\rotate(\k**)\rmov*(\d*,0pt){\sl*}}\relax
\advance\k**\r*
\advance\N*-1\ifnum\N*>0\repeat\advance\k**\m**
{\rotate(\k**)\mov(#2,0){\sl*}}{\rotate(#4)\mov(#2,0){\sl*}}}}

\def\gmov*#1(#2,#3)#4{\rlap{\L*=#1\Lengthunit
\xL*=#2\L* \yL*=#3\L*
\rx* \gcos*\xL* \tmp* \gsin*\yL* \advance\rx*-\tmp*
\ry* \gcos*\yL* \tmp* \gsin*\xL* \advance\ry*\tmp*
\rx*=\xscale\rx* \ry*=\yscale\ry*
\xL* \the\cos*\rx* \tmp* \the\sin*\ry* \advance\xL*-\tmp*
\yL* \the\cos*\ry* \tmp* \the\sin*\rx* \advance\yL*\tmp*
\kern\xL*\raise\yL*\hbox{#4}}}

\def\rgmov*(#1,#2)#3{\rlap{\xL*#1\yL*#2\relax
\rx* \gcos*\xL* \tmp* \gsin*\yL* \advance\rx*-\tmp*
\ry* \gcos*\yL* \tmp* \gsin*\xL* \advance\ry*\tmp*
\rx*=\xscale\rx* \ry*=\yscale\ry*
\xL* \the\cos*\rx* \tmp* \the\sin*\ry* \advance\xL*-\tmp*
\yL* \the\cos*\ry* \tmp* \the\sin*\rx* \advance\yL*\tmp*
\kern\xL*\raise\yL*\hbox{#3}}}

\def\Earc(#1)[#2,#3][#4,#5]{{\k*=#2\l*=#3\m*=\l*
\advance\m*-6\ifnum\k*>\l*\relax\else\def\xscale{#4}\def\yscale{#5}\relax
{\angle**0\rotate(#2)}\gmov*(#1,0){\sm*}\loop
\ifnum\k*<\m*\advance\k*5\relax
{\angle**0\rotate(\k*)}\gmov*(#1,0){\sl*}\repeat
{\angle**0\rotate(#3)}\gmov*(#1,0){\sl*}\relax
\def\xscale{1}\def\yscale{1}\fi}}

\def\dashEarc(#1)[#2,#3][#4,#5]{{\k**=#2\n*=#3\advance\n*-1\advance\n*-\k**
\L*=1000sp\L*#1\L* \multiply\L*\n* \multiply\L*\Nhalfperiods
\divide\L*57\N*\L* \divide\N*2000\ifnum\N*=0\N*1\fi
\r*\n*  \divide\r*\N* \ifnum\r*<2\r*2\fi
\m**\r* \divide\m**2 \l**\r* \advance\l**-\m** \N*\n* \divide\N*\r*
\k**\r*\multiply\k**\N* \dn*\n* \advance\dn*-\k** \divide\dn*2\advance\dn*\*one
\r*\l** \divide\r*2\advance\dn*\r* \advance\N*-2\k**#2\relax
\ifnum\l**<6\def\xscale{#4}\def\yscale{#5}\relax
{\angle**0\rotate(#2)}\gmov*(#1,0){\sm*}\advance\k**\dn*
{\angle**0\rotate(\k**)}\gmov*(#1,0){\sl*}\advance\k**\m**
{\angle**0\rotate(\k**)}\gmov*(#1,0){\sm*}\loop
\advance\k**\l**{\angle**0\rotate(\k**)}\gmov*(#1,0){\sl*}\advance\k**\m**
{\angle**0\rotate(\k**)}\gmov*(#1,0){\sm*}\advance\N*-1\ifnum\N*>0\repeat
{\angle**0\rotate(#3)}\gmov*(#1,0){\sl*}\def\xscale{1}\def\yscale{1}\else
\advance\k**\dn* \Earc(#1)[#2,\k**][#4,#5]\loop\advance\k**\m** \r*\k**
\advance\k**\l** {\Earc(#1)[\r*,\k**][#4,#5]}\relax
\advance\N*-1\ifnum\N*>0\repeat
\advance\k**\m**\Earc(#1)[\k**,#3][#4,#5]\fi}}

\def\triangEarc#1(#2)[#3,#4][#5,#6]{{\k**=#3\n*=#4\advance\n*-\k**
\L*=1000sp\L*#2\L* \multiply\L*\n* \multiply\L*\Nhalfperiods
\divide\L*57\N*\L* \divide\N*1000\ifnum\N*=0\N*1\fi
\d**=#2\Lengthunit \d*\d** \divide\d*57\multiply\d*\n*
\r*\n*  \divide\r*\N* \ifnum\r*<2\r*2\fi
\m**\r* \divide\m**2 \l**\r* \advance\l**-\m** \N*\n* \divide\N*\r*
\dt*\d* \divide\dt*\N* \dt*.5\dt* \dt*#1\dt*
\divide\dt*1000\multiply\dt*\magnitude
\k**\r* \multiply\k**\N* \dn*\n* \advance\dn*-\k** \divide\dn*2\relax
\r*\l** \divide\r*2\advance\dn*\r* \advance\N*-1\k**#3\relax
\def\xscale{#5}\def\yscale{#6}\relax
{\angle**0\rotate(#3)}\gmov*(#2,0){\sm*}\advance\k**\dn*
{\angle**0\rotate(\k**)}\gmov*(#2,0){\sl*}\advance\k**-\m**
\advance\l**\m**\loop\dt*-\dt* \d*\d** \advance\d*\dt*
\advance\k**\l**{\angle**0\rotate(\k**)}\rgmov*(\d*,0pt){\sl*}\relax
\advance\N*-1\ifnum\N*>0\repeat\advance\k**\m**
{\angle**0\rotate(\k**)}\gmov*(#2,0){\sl*}\relax
{\angle**0\rotate(#4)}\gmov*(#2,0){\sl*}\def\xscale{1}\def\yscale{1}}}

\def\waveEarc#1(#2)[#3,#4][#5,#6]{{\k**=#3\n*=#4\advance\n*-\k**
\L*=4000sp\L*#2\L* \multiply\L*\n* \multiply\L*\Nhalfperiods
\divide\L*57\N*\L* \divide\N*1000\ifnum\N*=0\N*1\fi
\d**=#2\Lengthunit \d*\d** \divide\d*57\multiply\d*\n*
\r*\n*  \divide\r*\N* \ifnum\r*=0\r*1\fi
\m**\r* \divide\m**2 \l**\r* \advance\l**-\m** \N*\n* \divide\N*\r*
\dt*\d* \divide\dt*\N* \dt*.7\dt* \dt*#1\dt*
\divide\dt*1000\multiply\dt*\magnitude
\k**\r* \multiply\k**\N* \dn*\n* \advance\dn*-\k** \divide\dn*2\relax
\divide\N*4\advance\N*-1\k**#3\def\xscale{#5}\def\yscale{#6}\relax
{\angle**0\rotate(#3)}\gmov*(#2,0){\sm*}\advance\k**\dn*
{\angle**0\rotate(\k**)}\gmov*(#2,0){\sl*}\advance\k**-\m**
\advance\l**\m**\loop\dt*-\dt*
\d*\d** \advance\d*\dt* \dd*\d** \advance\dd*1.3\dt*
\advance\k**\r*{\angle**0\rotate(\k**)}\rgmov*(\d*,0pt){\sl*}\relax
\advance\k**\r*{\angle**0\rotate(\k**)}\rgmov*(\dd*,0pt){\sl*}\relax
\advance\k**\r*{\angle**0\rotate(\k**)}\rgmov*(\d*,0pt){\sl*}\relax
\advance\k**\r*
\advance\N*-1\ifnum\N*>0\repeat\advance\k**\m**
{\angle**0\rotate(\k**)}\gmov*(#2,0){\sl*}\relax
{\angle**0\rotate(#4)}\gmov*(#2,0){\sl*}\def\xscale{1}\def\yscale{1}}}

\newcount\CatcodeOfAtSign
\CatcodeOfAtSign=\the\catcode`\@
\catcode`\@=11
\def\@arc#1[#2][#3]{\rlap{\Lengthunit=#1\Lengthunit
\sm*\l*arc(#2.1914,#3.0381)[#2][#3]\relax
\mov(#2.1914,#3.0381){\l*arc(#2.1622,#3.1084)[#2][#3]}\relax
\mov(#2.3536,#3.1465){\l*arc(#2.1084,#3.1622)[#2][#3]}\relax
\mov(#2.4619,#3.3086){\l*arc(#2.0381,#3.1914)[#2][#3]}}}

\def\dash@arc#1[#2][#3]{\rlap{\Lengthunit=#1\Lengthunit
\d*arc(#2.1914,#3.0381)[#2][#3]\relax
\mov(#2.1914,#3.0381){\d*arc(#2.1622,#3.1084)[#2][#3]}\relax
\mov(#2.3536,#3.1465){\d*arc(#2.1084,#3.1622)[#2][#3]}\relax
\mov(#2.4619,#3.3086){\d*arc(#2.0381,#3.1914)[#2][#3]}}}

\def\wave@arc#1[#2][#3]{\rlap{\Lengthunit=#1\Lengthunit
\w*lin(#2.1914,#3.0381)\relax
\mov(#2.1914,#3.0381){\w*lin(#2.1622,#3.1084)}\relax
\mov(#2.3536,#3.1465){\w*lin(#2.1084,#3.1622)}\relax
\mov(#2.4619,#3.3086){\w*lin(#2.0381,#3.1914)}}}

\def\bezier#1(#2,#3)(#4,#5)(#6,#7){\N*#1\l*\N* \advance\l*\*one
\d* #4\Lengthunit \advance\d* -#2\Lengthunit \multiply\d* \*two
\b* #6\Lengthunit \advance\b* -#2\Lengthunit
\advance\b*-\d* \divide\b*\N*
\d** #5\Lengthunit \advance\d** -#3\Lengthunit \multiply\d** \*two
\b** #7\Lengthunit \advance\b** -#3\Lengthunit
\advance\b** -\d** \divide\b**\N*
\mov(#2,#3){\sm*{\loop\ifnum\m*<\l*
\a*\m*\b* \advance\a*\d* \divide\a*\N* \multiply\a*\m*
\a**\m*\b** \advance\a**\d** \divide\a**\N* \multiply\a**\m*
\rmov*(\a*,\a**){\unhcopy\spl*}\advance\m*\*one\repeat}}}

\catcode`\*=12

\newcount\n@ast

\def\n@ast@#1{\n@ast0\relax\get@ast@#1\end}
\def\get@ast@#1{\ifx#1\end\let\next\relax\else
\ifx#1*\advance\n@ast1\fi\let\next\get@ast@\fi\next}

\newif\if@up \newif\if@dwn
\def\up@down@#1{\@upfalse\@dwnfalse
\if#1u\@uptrue\fi\if#1U\@uptrue\fi\if#1+\@uptrue\fi
\if#1d\@dwntrue\fi\if#1D\@dwntrue\fi\if#1-\@dwntrue\fi}

\def\halfcirc#1(#2)[#3]{{\Lengthunit=#2\Lengthunit\up@down@{#3}\relax
\if@up\mov(0,.5){\@arc[-][-]\@arc[+][-]}\fi
\if@dwn\mov(0,-.5){\@arc[-][+]\@arc[+][+]}\fi
\def\lft{\mov(0,.5){\@arc[-][-]}\mov(0,-.5){\@arc[-][+]}}\relax
\def\rght{\mov(0,.5){\@arc[+][-]}\mov(0,-.5){\@arc[+][+]}}\relax
\if#3l\lft\fi\if#3L\lft\fi\if#3r\rght\fi\if#3R\rght\fi
\n@ast@{#1}\relax
\ifnum\n@ast>0\if@up\shade[+]\fi\if@dwn\shade[-]\fi\fi
\ifnum\n@ast>1\if@up\dshade[+]\fi\if@dwn\dshade[-]\fi\fi}}

\def\halfdashcirc(#1)[#2]{{\Lengthunit=#1\Lengthunit\up@down@{#2}\relax
\if@up\mov(0,.5){\dash@arc[-][-]\dash@arc[+][-]}\fi
\if@dwn\mov(0,-.5){\dash@arc[-][+]\dash@arc[+][+]}\fi
\def\lft{\mov(0,.5){\dash@arc[-][-]}\mov(0,-.5){\dash@arc[-][+]}}\relax
\def\rght{\mov(0,.5){\dash@arc[+][-]}\mov(0,-.5){\dash@arc[+][+]}}\relax
\if#2l\lft\fi\if#2L\lft\fi\if#2r\rght\fi\if#2R\rght\fi}}

\def\halfwavecirc(#1)[#2]{{\Lengthunit=#1\Lengthunit\up@down@{#2}\relax
\if@up\mov(0,.5){\wave@arc[-][-]\wave@arc[+][-]}\fi
\if@dwn\mov(0,-.5){\wave@arc[-][+]\wave@arc[+][+]}\fi
\def\lft{\mov(0,.5){\wave@arc[-][-]}\mov(0,-.5){\wave@arc[-][+]}}\relax
\def\rght{\mov(0,.5){\wave@arc[+][-]}\mov(0,-.5){\wave@arc[+][+]}}\relax
\if#2l\lft\fi\if#2L\lft\fi\if#2r\rght\fi\if#2R\rght\fi}}

\catcode`\*=11

\def\Circle#1(#2){\halfcirc#1(#2)[u]\halfcirc#1(#2)[d]\n@ast@{#1}\relax
\ifnum\n@ast>0\L*=\xscale\Lengthunit
\ifnum\angle**=0\clap{\vrule width#2\L* height.1pt}\else
\L*=#2\L*\L*=.5\L*\special{em:linewidth .001pt}\relax
\rmov*(-\L*,0pt){\sm*}\rmov*(\L*,0pt){\sl*}\relax
\special{em:linewidth \the\linwid*}\fi\fi}

\catcode`\*=12

\def\wavecirc(#1){\halfwavecirc(#1)[u]\halfwavecirc(#1)[d]}
\def\dashcirc(#1){\halfdashcirc(#1)[u]\halfdashcirc(#1)[d]}

\def\xscale{1}

\def\yscale{1}

\def\Ellipse#1(#2)[#3,#4]{\def\xscale{#3}\def\yscale{#4}\relax
\Circle#1(#2)\def\xscale{1}\def\yscale{1}}

\def\dashEllipse(#1)[#2,#3]{\def\xscale{#2}\def\yscale{#3}\relax
\dashcirc(#1)\def\xscale{1}\def\yscale{1}}

\def\waveEllipse(#1)[#2,#3]{\def\xscale{#2}\def\yscale{#3}\relax
\wavecirc(#1)\def\xscale{1}\def\yscale{1}}

\def\halfEllipse#1(#2)[#3][#4,#5]{\def\xscale{#4}\def\yscale{#5}\relax
\halfcirc#1(#2)[#3]\def\xscale{1}\def\yscale{1}}

\def\halfdashEllipse(#1)[#2][#3,#4]{\def\xscale{#3}\def\yscale{#4}\relax
\halfdashcirc(#1)[#2]\def\xscale{1}\def\yscale{1}}

\def\halfwaveEllipse(#1)[#2][#3,#4]{\def\xscale{#3}\def\yscale{#4}\relax
\halfwavecirc(#1)[#2]\def\xscale{1}\def\yscale{1}}

\catcode`\@=\the\CatcodeOfAtSign

\title{\boldmath Supersymmetric galileons and auxiliary fields in 2+1 dimensions}
\preprint{KA-TP-44-2016}
\author{Jose M. Queiruga }
\affiliation{Institute for Theoretical Physics, Karlsruhe Institute
of Technology (KIT), 76131 Karlsruhe, Germany}
\email{jose.queiruga@kit.edu}

\begin{abstract}
In this work we study various aspects of supersymmetric three-dimensional higher-derivative field theories. We classify all possible models without derivative terms of the auxiliary field in the fermionic sector and find that scalar field theories of the form $P(X,\phi)$, where $X=-(\pa\phi)^2/2$, belong to this kind of models. A ghost-free supersymmetric extension of Galileon models is found in three spacetime dimensions. Finally, the auxiliary field problem is discussed. 
\end{abstract}

\maketitle


\section{Introduction}
\label{sec:intro}

Higher-derivative field theories (the so-called K-theories) have been increasing in importance in recent years. It has been understood that nonstandard derivative terms (higher powers of the first order derivatives as well as higher derivative terms) can emerge in effective field theories in a natural and a rather unavoidable way. As the most famous example of a classical field theory with a higher-derivative term let us mention 
the Skyrme model \cite{Skyrme1}, where the usual quadratic term in derivatives (Dirichlet term) must be accompanied by a quartic term (the Skyrme term) or by a sextic term. The Skyrme model has been applied successfully in the study of atomic nuclei and nuclear matter \cite{ANW}. It plays also an important role in the holographic study of QCD \cite{Sakai}. Furthermore, its low-dimensional analogues have been applied to condensed matter \cite{Sondhi, Schwindt} or cosmology \cite{Kodama, Brihaye, Delsate}.
More recently phenomena like K-inflation \cite{Armendariz1} or K-essence \cite{Armendariz2} have been described in terms of higher-derivative field theories. K-theories are widely used in cosmology \cite{Babichev1, Babichev2, Olechowski, Andrews}, DBI inflation \cite{Silverstein}, Ghost-condensates \cite{Arkani} and compacton models \cite{weres}. 
The last example of such higher-derivative models is provided by Galileon theories \cite{Deffayet, Luty, Nicolis1, Nicolis2, Chow, Agarwal, Hinterbichler, Andrews2}, where,  despite the fact that higher-derivative terms are present in the action, the equations of motion are at most of second order in derivatives.

Since K-theories can be considered as effective field theories arising at a certain limit of more fundamental ones (see for example \cite{Sakai} and \cite{Rham}) one may ask about the fate of the supersymmetry. Namely, if some of the supersymmetry of the fundamental theory is preserved when such a limit is taken it is natural to consider these effective models in a supersymmetric context.  For some years the supersymmetric extensions of these models have attracted the attention of theoretical physicists, namely, SUSY Skyrme-like models \cite{Nepomechie, Freyhult, q1, q2, q3, q4, Nitta1}, SUSY Ghost-condensates \cite{Khoury1, Koehn1}, SUSY Galileons in four dimensions \cite{Khoury2, Koehn2, Farakos}, SUSY k-defects \cite{Koehn3, q5, q6} and general higher-derivative field theories \cite{Nitta3, Nitta4, Nitta5, Antoniadis, Koehn4, Petrov}.

When higher-derivative terms are present in the action, the theories are frequently accompanied by some ``pathology". There are two problems which usually arise in this context: the ghost problem and the appearance of dynamical auxiliary fields.  

Due to the Ostrogradski theorem (see \cite{Woodard1,Woodard2}) one can expect a possible appearance of ghost degrees of freedom. However, certain types of K-theories are free from this problem. The principal examples are contained in the following categories: the so-called Skyrme like models, Ghost-condensates \cite{Arkani} and Galileon theories \cite{Deffayet, Luty, Nicolis1, Nicolis2, Chow, Agarwal, Hinterbichler, Andrews2}. In the case of SUSY Galileon theories it has been found that the cubic Galileon contains ghost-like states \cite{Koehn2}. This unwanted feature can be cured for the quartic Galileon and a ghost-free SUSY version can be constructed in this case \cite{Farakos}.

Another of the peculiarities in SUSY theories is the necessary presence of auxiliary fields $F$ (in the off-shell action), which normally do not correspond to physical degrees of freedom (d.o.f) and can be eliminated algebraically. However, as it was pointed out long ago in \cite{Nepomechie}, although $F$ shows up algebraically in the bosonic sector, the fermionic sector may contain derivatives acting on $F$ \cite{Gates1, Gates2, Komargodski, Nitta6}. Let us remark that the fact that $F$ can become dynamical is not necessarily a problem. In this case the spinor in the supermultiplet must have an extra fermionic d.o.f to balance the bosonic ones (in three dimensions and $N=1$ a real superfield contains usually one fermionic and one bosonic). This extra d.o.f. may be of a ghost-like type. In some situations these potentially dangerous terms containing derivative acting on $F$ can be eliminated from the action by adding some extra terms \cite{Khoury2,Nitta6}. Recently it has been shown that it is possible to construct a ghost-free SUSY action with propagating auxiliary fields \cite{Nitta7}. 

This is the main goal of the present paper: to analyze these two issues in a more systematic way. Specifically, we want to understand under which circumstances non-standard derivative field theories in three spacetime dimensions (K-models, galileons) lead to supersymmetric extensions with a non-dynamical auxiliary field. 

The outline of the paper is as follows. In Sec. II we present several families of supersymmetric actions containing non-trivial bosonic sector to be widely used in the next sections. In Sec. III we restrict these families of theories to those without derivatives acting on the auxiliary field. In Sec. IV  Galileon theories in three dimensions are supersymmetrized. We show that in three dimensions all  SUSY Galileon theories are ghost-free (see \cite{Koehn2} for a counterexample in four dimensions). Furthermore, these supersymmetric forms do not contain derivatives acting on $F$ (in four dimensions this kind of terms seems to be unavoidable \cite{Farakos}). We also show a general result about the existence of supersymmetric extensions of scalar theories. In Sec. V we explore the consequence of the presence of $F$-derivative terms. We show that when these terms appear in the fermionic sector, $F$ can be always eliminated algebraically. In Sec. V.A we discuss the relation between F-derivative terms in the bosonic sector and ghost states. Finally Sec. VI is devoted to our summary. The appendices include useful D-algebra identities and some considerations about the explicit solution for $F$.


\section{supersymmetric higher derivative actions in 3 dimensions}

The most general $N=1$ supersymmetric action in three dimensions involving one real superfield can be written as follows
\be
\mathcal{L}=\sum_i \alpha_i\int d^2\theta\, \mathcal{O}_1\Phi...\mathcal{O}_n\Phi\label{genact}
\ee
where $\Phi$ is a real scalar superfield 
\be
\Phi=\phi+\theta^\alpha\psi_\alpha-\theta^2 F\label{rfield}
\ee
and the operators $\mathcal{O}_i$ involve $n_i$ superderivatives
\be
\mathcal{O}_i= D D ...\overset{n_i}{...}...D\label{ope}.
\ee

The spinor indices of the superderivatives can be contracted in different ways but we do not display them in (\ref{ope}) for the sake of notational simplicity. The superderivative is defined by
\be
D_\alpha=\pa_\alpha+i\theta^\beta \pa_{\alpha\beta}, \quad \pa_\alpha\equiv\frac{\pa}{\pa \theta^\alpha}\quad\text{and}\quad \pa_{\alpha\beta}\equiv\sigma_{\alpha\beta}^\mu\pa_\mu\label{superd}.
\ee

Besides all possible non-commuting operators constructed in terms of $D$-operators, it is possible to build commuting ones. Specifically, the number operator corresponding to $n_i=0$ in (\ref{ope}), or spacetime derivatives, which can be built in terms of anticommutators of $D$-operators,
\be
\lbrace D_\alpha,D_\beta\rbrace=2i\pa_{\alpha\beta}.
\ee

We define the order of $\mathcal{O}_i$ as the number of superderivatives present in (\ref{ope})
\be
\lbrack \mathcal{O}_i\rbrack=n_i.
\ee

We can distinguish two different types of operators. First, if $\lbrack \mathcal{O}_i\rbrack$ is an even number, the operator $\mathcal{O}_i$ is bosonic (the component expansion of $\mathcal{O}_i \Phi$ setting $\theta=0$ does not involve fermions). Second, if $\lbrack \mathcal{O}_i\rbrack$ is an odd number, the operator $\mathcal{O}_i$ is  fermionic, and its component expansion only involves fermions. We define accordingly the degree of $\mathcal{O}_i$ as follows
\bea
&\text{if}&\quad \lbrack \mathcal{O}_i\rbrack\in 2\mathbb{Z}:\Rightarrow \text{deg}\mathcal{O}_i=0\\
&\text{if}&\quad \lbrack \mathcal{O}_i\rbrack\in 2\mathbb{Z}+1:\Rightarrow \text{deg}\mathcal{O}_i=1.
\eea

We define the degree of a product of operators as follows
\be
\text{deg}\left(\Pi_i \mathcal{O}_i\Phi\right)=\Pi_i\left(1-\text{deg}\mathcal{O}_i\right)+1 \left(\text{mod 2}\right)\label{deg}.
\ee

The degree of the product distinguishes between two situations, namely, if the component expansion of $\Pi_i \mathcal{O}_i\Phi$ (setting $\theta=0$)  contains or does not contain fermions (respec. $\text{deg}\left(\Pi_i \mathcal{O}_i\Phi\right)=1$ or $\text{deg}\left(\Pi_i \mathcal{O}_i\Phi\right)=0$). We can now study the existence of non-trivial bosonic sector in three dimensional supersymmetric Lagragians. Let us assume for simplicity that the Lagrangian (\ref{genact}) consists of one single term (the subsequent analysis applies trivially for the sum), 
\be
\mathcal{L}=\int d^2\theta\, \mathcal{O}_1 \Phi...\mathcal{O}_n\Phi\label{act1}.
\ee 

Due to the properties of the Grassmann integration (\ref{gras}) we can expand (\ref{act1}) in terms of superderivatives
\bea
\mathcal{L}&=&D^2\mathcal{O}_1\Phi...\mathcal{O}_n\Phi+...+\mathcal{O}_1\Phi...D^2\mathcal{O}_n\Phi-\frac{1}{2}D^\alpha \mathcal{O}_1\Phi D_\alpha\mathcal{O}_2\Phi...\mathcal{O}_n\Phi-...\nonumber\\
&-&\frac{1}{2}D^\alpha \mathcal{O}_1\Phi \mathcal{O}_2\Phi...D_\alpha\mathcal{O}_n+...+\frac{1}{2}\mathcal{O}_1...D^\alpha \mathcal{O}_{n-1}D_\alpha\mathcal{O}_n\Phi\vert\label{actexp}
\eea
where $\vert$ means that we set all $\theta$'s to zero after differentiation. The degree has the following property
\bea
\text{deg}\left(D^2\mathcal{O}_i\right)&=&\text{deg}\mathcal{O}_i\label{id1}\\
\text{deg}\left(D_\alpha\mathcal{O}_i\right)&=&\text{deg}\mathcal{O}_i+1\left(\text{mod 2}\right).\label{id2}
\eea

If we set
\bea
a_1&=&D^2\mathcal{O}_1\Phi...\mathcal{O}_n\Phi\vert\\
a_2&=&\mathcal{O}_1\Phi D^2 \mathcal{O}_2...\mathcal{O}_n\Phi\vert\\
&...&\nonumber\\
a_n&=&\mathcal{O}_1\Phi...D^2\mathcal{O}_n\Phi\vert\\
&...&\nonumber\\
a_{\frac{1}{2}n(n+1)}&=&\mathcal{O}_1\Phi...D^\alpha \mathcal{O}_{n-1}D_\alpha\mathcal{O}_n\Phi\vert
\eea
then
\be
\mathcal{L}=\sum_{i\in I}a_i.
\ee

The bosonic sector will be
\be
\mathcal{L}\vert_{\text{bos}}=\sum_{j\in J\subset I}a_j,\quad\text{such that}\quad \text{deg}\,a_j=0.
\ee

From here we draw two obvious consequences: 1) or $J=\emptyset$ and the bosonic sector is trivial ($=0$) or 2)  $J\neq I$ and since in the expansion (\ref{actexp}) or $\text{deg}\left(D^2\mathcal{O}_i\right)=1$ or $\text{deg}\left(D^\alpha\mathcal{O}_i\right)=1$, there are always fermionic terms in the expansion. We can now classify all possible supersymmetric Lagrangians with non-trivial bosonic sector. They split into two different types. The canonical form of the first type is
\be
\mathcal{L}^{\text{type I}}=\int d^2\theta \Pi_{i\in I} \mathcal{O}_i\Phi,\quad\text{such that  }\text{deg}\left(\Pi_{i\in I} \mathcal{O}_i\Phi \right)=0.
\ee

According to the definition (\ref{deg}) we have the following
\be
 \text{deg}\left(\Pi_{i\in I} \mathcal{O}_i\Phi \right)=0\Rightarrow \text{deg}\mathcal{O}_i=0,\,\forall\,i\in I\label{tyI}.
\ee

In particular the bosonic sector is given by
\be
\mathcal{L}^{\text{type I}}\vert_{\text{bos}}=D^2 \mathcal{O}_1 \Phi...\mathcal{O}_n\Phi+...+ \mathcal{O}_1\Phi...D^2\mathcal{O}_n\Phi\vert
\ee
(we see from (\ref{id1}) that the sum above only contains bosonic operators). The second type of Lagrangian has the form
\be
\mathcal{L}^{\text{type II}}=\int d^2\theta \Pi_{i\in I} \mathcal{O}_i\Phi\,\,\text{such that  }\overset{..}{\exists}i,j\in I\,\text{deg} \mathcal{O}_i=\text{deg} \mathcal{O}_j=1
\ee
i.e. $\forall k\in I, k\neq i,j \Rightarrow \text{deg}\mathcal{O}_k=0$. In this case the bosonic sector contains only one term
\be
\mathcal{L}^{\text{type II}}\vert_\text{bos}= \mathcal{O}_1\Phi...D^\alpha\mathcal{O}_i\Phi...D_\alpha \mathcal{O}_j\Phi... \mathcal{O}_n\Phi.
\ee

The simplest example of type I Lagrangians is the prepotential $W(\Phi)$. If we expand $W(\Phi)$ in powers of $\Phi$ we get
\be
W(\Phi)=\sum_i\alpha_i \Phi^i\Rightarrow \int d^2 \theta W(\Phi)=\sum_i \alpha_i\int d^2\theta \Phi^i\label{ty1}.
\ee

The sum in the right-hand side of (\ref{ty1}) is of the form (\ref{tyI}) where $\mathcal{O}_i=1$. The simplest example of type II Lagrangians is the supersymmetric non-linear sigma model
\be
 g(\Phi)D^\alpha \Phi D_\alpha\Phi=\sum_i\beta_i\Phi^i D^\alpha \Phi D_\alpha\Phi.
\ee

The pair of odd operators are $\mathcal{O}_i=D^\alpha$ and $\mathcal{O}_j=D_\alpha$. For $k\neq i,j$ the operators are c-numbers and therefore $\deg \mathcal{O}_k\Phi=0$. A remark is in order: as long as type I Lagrangians have at least one $D$-operator, they are related with type II Lagrangians up to fermionic terms.  This can be seen trivially integrating by parts. For simplicity we have considered single superfield theories, but these results can be trivially extended to models with various superfields. We emphasize again the fact there are no more possibilities to generate Lagrangians with non-trivial bosonic sector.  For example, the presence of exactly one or more than two odd operators leads to a purely fermionic model. This fact has an interesting consequence. If we have a SUSY extension of a given bosonic model, any deformation of the model consisting of terms not belonging to type I or type II families ($=$pure fermionic terms) does not change the bosonic sector. For example, the canonical supersymmetric form of the linear $\sigma$-model with potential is given by
\be
\mathcal{L}=\int d^2\theta \left(D^\alpha \Phi D_\alpha \Phi+W(\Phi)\right).\label{sigma}
\ee

We can deform this model with a pure fermionic term of the form
\be
\mathcal{L}_d=\lambda\int d^2\theta D^\alpha \Phi D_\alpha \Phi D^2 D^\beta \Phi D^2D_\beta \Phi.\label{defo}
\ee

The resulting model ($\mathcal{L}+\mathcal{L}_d$) has the same bosonic sector of  (\ref{sigma}), in components
\bea
\mathcal{L}+\mathcal{L}_d&=&\left(-\pa_\mu\phi\pa^\mu\phi+i\psi^\alpha\pa_\alpha^{\,\,\,\beta}\psi_\beta+F^2+W'(\phi)F+W''(\phi)\psi^\alpha\psi_\alpha\right)\nonumber\\
&-&\lambda\left(2 i \pa^{\alpha}_{\,\,\,\beta}\psi^\beta \psi_\alpha+\pa^{\gamma\alpha}\phi\pa_{\gamma\alpha}\phi-2 F^2\right) \pa^{\beta}_{\,\,\,\alpha}\psi^\alpha\pa_{\beta\gamma}\psi^\gamma\nonumber\\
&+&\lambda\left(2i\square \psi^\beta\pa_{\alpha\beta}\psi^\beta-\pa^{\alpha\beta}F\pa_{\alpha\beta}F-2(\square\phi)^2\right)\psi^\alpha\psi_\alpha\nonumber\\
&+&4\lambda\left(i\pa^{\gamma\alpha}\phi\pa_\gamma^{\,\,\, \beta} F-\pa^{\beta\alpha}\phi\square\phi+\pa^{\alpha\beta}F F-i F\square\phi C^{\beta\alpha}\right)\psi_\alpha\pa_{\beta	\delta}\psi^\delta\label{susyl1}
\eea


and therefore
\be
\mathcal{L}\vert_{\psi=0}=\mathcal{L}+\mathcal{L}_d\vert_{\psi=0}.
\ee

 We can multiply any fermionic term with an arbitrary analytic function of the superfields  without changing its fermionic character (for example $\lambda\int d^2\theta \, h(\Phi) D^\alpha \Phi D^2D_\alpha \Phi D^\beta \Phi D^2 D_\beta \Phi$) leading to an infinite number of SUSY extensions. We call these SUSY models sharing the same bosonic sector ``SUSY bosonic twins". We can formulate this result in the following lemma:

   \begin{lemma}
\label{infextensions1}
If a three dimensional bosonic scalar model possesses an $N=1$ SUSY extension then it possesses infinitely many of them.
\end{lemma}

The fermionic deformation in (\ref{susyl1}) leads to higher-derivative terms in the fermionic sector. This is because we have introduced the operators $D^2 D_\alpha$ in the superaction. These higher derivative terms can be avoided for example if instead of (\ref{defo}), we introduce $ D^\alpha \Phi D_\alpha \Phi  D^\beta \tilde{\Phi} D_\beta \tilde{\Phi}$, being $\tilde{\Phi}$ another real superfield (note that when $\Phi=\tilde{\Phi}$ this term is identically zero). Other interesting feature of (\ref{susyl1}) is the appearance of terms with derivatives for the auxiliary field, but as we will see later, this does not imply that $F$ becomes dynamical.

\section{General models and the auxiliary fields}

In the previous section we have classified all possible supersymmetric actions with non-trivial bosonic sector. These actions can contain in general higher-derivative terms (in the scalar field $\phi$), but also derivatives acting on the auxiliary field $F$. The presence of such terms leads in general to propagating auxiliary fields (we will discuss this point in more detail later), and therefore the equation of motion for $F$ could be non-algebraic. From the superfield expression it is easy to see that terms of the form $\pa F$ are generated in the bosonic sector when an operator of the form $\pa D^2$ acts on $\Phi$ after the Grassmann integration. On the other hand, the Grassmann integration is equivalent to the supersymmetric differentiation under the replacement
\be
\int d^2\theta \,\Omega \rightarrow D^2 \Omega\vert\label{gras}
\ee
and therefore it increases the order or the operators $\mathcal{O}_i$ at most by 2.
 The first operator $\pa D^2$ appears at order 4 (see Appendix A) unless the operator is of the form $D^2$ (\ref{idap5}). We arrive at the following result: the most general Lagrangian of type I not involving F-derivatives in the bosonic sector is of the form
 \be
 \mathcal{L}^{\text{type I }\star}=\int d^2\theta\, P(\Phi,D^2\Phi,D^\alpha D^\beta\Phi, \pa_{\alpha\beta}\Phi)\label{type1s}.
 \ee
 
 This kind of actions can generate linear terms in $\pa F$ but they can always be eliminated after integration by parts (see the example below). A similar argument restricts the type II Lagrangians to the form 
  \be
 \mathcal{L}^{\text{type II }\star}=\int d^2\theta\, D_\alpha\Phi \mathcal{O}_\beta \Phi \left(\Pi_i \mathcal{O}_i\right)^{\alpha\beta},  \quad [\mathcal{O}_\beta]\in\{1,3\}, \quad [\mathcal{O}_i] \in\{0,2\}\,\text{or}\,\mathcal{O}_i=f(\pa).\label{type2s}
 \ee
where $f(\pa)$ stands for any operator involving only space-time derivatives. It is important to note that $[\pa]=2$ since $2i\pa_{\alpha\beta}=\lbrace D_\alpha,D_\beta \rbrace$. Moreover, it is easy to see that $ \mathcal{L}^{\text{type I }\star}$ cannot contain F-derivatives in the fermionic sector. If we want to avoid the proliferation of these terms in the fermionic sector for    $\mathcal{L}^{\text{type II }\star}$ we need an extra constraint on the $\mathcal{O}_i$ operators
 \be
 \mathcal{L}^{\text{type II }\star\star}=\int d^2\theta\, D_\alpha\Phi \mathcal{O}_\beta \Phi \left(\Pi_i \mathcal{O}_i\right)^{\alpha\beta},  \quad [\mathcal{O}_\beta]\in\{1,3\}, \quad [\mathcal{O}_i] \in\{0,2\}.\label{condII}.
 \ee

Therefore we do not allow the operators $\mathcal{O}_i$ to be higher-derivative terms (see the example in Sect. V). The nonlinear $\sigma$-model term is again an example of this family. A less trivial example of the supersymmetric extension of type II$\star\star$ models are $P(X)$-theories in three dimensions \cite{q3} (scalar field theories whose kinetic terms are analytic functions of $X=-\pa_\mu\phi\pa^\mu\phi$). If we define
\be
\mathcal{L}^{(k,n)}=-\frac{1}{2}\int d^2\theta D^\alpha \Phi D_\alpha\Phi \left(D^\beta D^\alpha\Phi D_\beta D_\alpha\Phi\right)^{k-1}\left(D^2\Phi D^2\Phi\right)^n 
\ee
we have
\be
\mathcal{L}^{(k)}\vert_{\psi=0}=\sum_{n=0}^{k-1}(-1)^{n}\left(\begin{matrix}
    k   \\
     n
\end{matrix}\right)\mathcal{L}^{(k-n,n)}\vert_{\psi=0}=X^k+(-1)^{k-1}F^{2k}.
\ee

Finally adding a prepotential
\be
\mathcal{L}=\sum_i \alpha_i \mathcal{L}^{(i)}+W'(\phi)F \vert_{\psi=0}=\sum_i\alpha_i\left(X^i+(-1)^{k-1}F^{2k}\right)+W'(\phi)F\label{PXt}.
\ee

If we choose the coefficients $\alpha_i$ such that $P(X)=\sum_i \alpha_i X^i$ the Lagrangian (\ref{PXt}) constitutes a supersymmetric extension of the $P(X)$-theory with potential (the choice $W(\phi)=0$ corresponds to the pure $P(X)$ since $F=0$ is always a solution of the equation of motion of the auxiliary field). Let us see what happens in the fermionic sector. The unique term potentially involving derivatives on F can be written as follows
\bea
&&(k-1)D^\alpha \Phi D_\alpha\Phi \left(D^\beta D^\alpha\Phi D_\beta D_\alpha\Phi\right)^{k-2}D^2 D^\beta D^\alpha\Phi D_\beta D_\alpha\Phi\vert=\nonumber\\
&&=(k-1)\psi^\alpha \psi_\alpha \left(-\pa_\mu\phi\pa^\mu \phi+F^2\right)^{k-2}\left(i\pa^{\beta\alpha}F+C^{\alpha\beta}\square\phi\right)\left(i\pa_{\beta\alpha}\phi+C_{\alpha\beta}F\right).
\eea

Now since $ C_{\alpha\beta}\pa^{\alpha\beta}=0$ holds, only survive one $\pa F$-term survives, namely
\be
(k-1)\psi^\alpha \psi_\alpha \left(-\pa_\nu\phi\pa^\nu \phi+F^2\right)^{k-2}\pa_\mu\phi\pa^\mu F
\ee 
which can be written up to a total derivative as
\be
(k-1)\sum_{i=0}^{k-2}\frac{1}{2i+1}\left(\begin{matrix}
    k-2   \\
     i
\end{matrix}\right)\pa^\mu \left(\left(-\pa_\nu\phi\pa^\nu\phi\right)^{k-2}\pa_\mu \phi\right)F^{2i+1}.
\ee

We have shown therefore that $F$ is also algebraic in the fermionic sector. In Sec. V we will return to the problem of terms $\pa F$ in the fermionic sector.


\section{Supersymmetric Galileons in three dimensions}

Galileon theories are higher-derivative field models, such that their equations of motion have at most two derivatives acting on the fields. The most general Galileon theory is a combination of the following Lagrangians
\bea
\mathcal{L}_{G,2}&=&-\frac{1}{2}\left(\pa \phi\right)^2\label{gal1}\\
\mathcal{L}_{G,3}&=&-\frac{1}{2}\left(\pa \phi\right)^2\square\phi\\
\mathcal{L}_{G,4}&=&\left(\pa \phi\right)^2\left(-\frac{1}{2}\left(\square\phi\right)^2+\frac{1}{2}\pa^\mu\pa^\nu\phi\pa_\mu\pa_\nu\phi\right)\label{gal4}.
\eea

In order to make explicit the galilean symmetry
\be
 \phi\rightarrow \phi+c+b_\mu x^\mu,\quad  \pa_\mu c=\pa_\mu b_\nu=0\label{galsym}
 \ee 
we can integrate by parts (\ref{gal1})-(\ref{gal4}) and rewrite the Lagrangians as 
\bea
\mathcal{L}_{G,2}&\propto&\epsilon^{\mu\rho\sigma}\epsilon^{\nu}_{\,\,\,\rho\sigma}\pa_\mu\phi\pa_\nu\phi\label{gall1}\\
\mathcal{L}_{G,3}&\propto&\epsilon^{\mu\rho\delta}\epsilon^{\nu\sigma}_{\,\,\,\,\,\,\delta}\pa_\mu\phi\pa_\nu\phi\pa_\rho\pa_\sigma\phi\\
\mathcal{L}_{G,4}&\propto&\epsilon^{\mu\rho\sigma}\epsilon^{\nu\delta\alpha}\pa_\mu\phi\pa_\nu\phi\pa_\rho\pa_\delta\phi\pa_\sigma\pa_\alpha\phi\label{gall4}.
\eea

In the light of (\ref{gall1})-(\ref{gall4}) it is clear that in three dimensions we cannot construct Lagrangians beyond (\ref{gal4}) respecting the Galilean symmetry. These theories can be supersymmetrized as type II models as follows
\bea
\mathcal{L}_{G,2}^S&=&\frac{1}{2}D^\alpha\Phi D_\alpha \Phi\label{ssgal1}\\
\mathcal{L}_{G,3}^S&=&\frac{1}{2}D^\alpha\Phi D_\alpha\Phi\square\Phi\label{ssgal2}\\
\mathcal{L}_{G,4}^S&=&-D^\alpha\Phi D_\alpha \Phi\left(-\frac{1}{2}\left(\square\Phi\right)^2+\frac{1}{2}\pa^\mu\pa^\nu\Phi\pa_\mu\pa_\nu\Phi\right)\label{ssgal4}.
\eea

The Galilean shift (\ref{galsym}) can be easily supersymmetrized:
\be
 \Phi\rightarrow \Phi+c+b_\mu x^\mu,\quad  \pa_\mu c=\pa_\mu b_\nu=0\label{sgalsym}.
 \ee 
 
In the $N=2$ case or in four dimensions this super-Galilean shift has an extra contribution (see \cite{Farakos}) since it is necessary to take into account the shift of the derivative of the field $\phi$ contained in the chiral superfield $\Phi\rightarrow\Phi+c+b_\mu(x^\mu+i\theta\sigma^\mu\bar{\theta})$. In three dimensions and $N=1$ SUSY the field $\phi$ is contained only in the lowest component of the real superfield (\ref{rfield}). We have to check now that the superfield Lagrangians are invariant under the super-Galilean shift (\ref{sgalsym}). Let us start with (\ref{ssgal1}). The transformation (\ref{sgalsym})  (see Appendix A) implies
\bea
\mathcal{L}_{G,2}^S&\rightarrow&\mathcal{L}_{G,2}^S+\int d^2\theta\left(2i\theta^\beta\sigma^\mu_{\beta\alpha}b_\mu D^\alpha\Phi-\theta^\beta\sigma^\mu_{\beta\alpha}b_\mu\theta^\gamma\sigma^{\nu\,\,\alpha}_\gamma b_\nu\right)=\nonumber\\
&=&\mathcal{L}_{G,2}^S+2b_\mu b^\mu+4 \pa_\mu b^\mu \pa_\mu\phi
\eea
where the last term is a total derivative after integration by parts. Now it is easy to see that (\ref{ssgal2}) and (\ref{ssgal4}) are also invariant under (\ref{sgalsym}) since $\square\Phi\rightarrow \square\Phi$ and $\pa_\mu\pa_\nu\Phi\rightarrow \pa_\mu\pa_\nu\Phi$. In terms of the component fields the bosonic sectors have the following form
\bea
\mathcal{L}_{G,2}^S\vert_{\psi=0}&=&\frac{1}{2}\left(-\left(\pa \phi\right)^2+F^2\right)\label{sgal1}\\
\mathcal{L}_{G,3}^S\vert_{\psi=0}&=&\frac{1}{2}\left(-\left(\pa \phi\right)^2+F^2\right)\square\phi\label{sgal2}\\
\mathcal{L}_{G,4}^S\vert_{\psi=0}&=&-\left(-\left(\pa \phi\right)^2+F^2\right) \left(-\frac{1}{2}\left(\square\phi\right)^2+\frac{1}{2}\pa^\mu\pa^\nu\phi\pa_\mu\pa_\nu\phi\right)\label{sgal4}.
\eea

Since the auxiliary field shows up quadratically in the bosonic sector we can take the solution $F=0$ and therefore the Lagrangians (\ref{sgal1})-(\ref{sgal4}) reduce to (\ref{gal1})-(\ref{gal4}). It is important to note that these supersymmetric Galileons are of type II$\star$ (not type II$\star\star$), except for the trivial case $\mathcal{L}_{G,2}^S$  and this implies in particular that terms with derivatives in $F$ will appear in the fermionic sector. The existence of this kind of terms does not imply necessarily that $F$ is a dynamical field as we will see below. Let us write down the full supersymmetric Lagrangian for $\mathcal{L}_{G,3}^S$
\be
\mathcal{L}_{G,3}^S=\frac{1}{2}\left(-(\pa_\mu\phi)^2+F^2+\psi^\alpha \pa_\alpha^{\,\,\beta}\psi_\beta\right)\square\phi-\frac{1}{2}\psi^\alpha\psi_\alpha\square F-i\pa^{\alpha\beta}\phi\psi_\alpha\square\psi_\beta-F\psi_\alpha\square\psi^\alpha.
\ee

The $F$-derivative term is linear in $F$, and therefore we can integrate by parts and remove the derivative. We obtain
\be
F=\frac{1}{2\square\phi}\pa_\mu\psi^\alpha\pa^\mu\psi_\alpha\label{auxgal3}.
\ee

We can immediately extend this result to the rest of the Galileon actions. This can be seen as follows: when we act with the $D^2$ operator on the $\sigma$-model term ($D^\alpha \Phi D_\alpha \Phi$) we generate a quadratic term in $F$ without derivatives. Now in the higher-derivative part of (\ref{ssgal2})-(\ref{ssgal4}) there are no $D^2$ operators, so the action of $D^2$ generates at most one term of the form $f(\pa)D^2$ per term, which corresponds in components to $f(\pa)F$. Since only one of these terms can appear, we can integrate by parts and transfer the derivatives to the rest of the fields. For $\mathcal{L}_{G,4}^S$ we get
\bea
\mathcal{L}_{G,4}^S&=&-\left(-(\pa_\mu\phi)^2+F^2+\psi^\alpha \pa_\alpha^{\,\,\beta}\psi_\beta\right)\left(-\frac{1}{2}\left(\square\phi\right)^2+\frac{1}{2}\pa^\mu\pa^\nu\phi\pa_\mu\pa_\nu\phi\right)\nonumber\\
&+&\psi^\alpha\psi_\alpha\left(-\frac{1}{2}\left(2\square F\square\phi+\square\psi^\alpha\square\psi_\alpha\right)+\frac{1}{2}\left(2\pa^\mu\pa^\nu F\pa_\mu\pa_\nu\phi+\pa^\mu\pa^\nu\psi^\beta\pa_\mu\pa_\nu\psi_\beta\right)\right)\nonumber\\
&+&2\left(i\pa^{\beta\alpha}\phi+C^{\alpha\beta}F\right)\psi_\alpha\left(-\square\psi_\beta\square\phi+\pa^\mu\pa^\nu\psi_\beta\pa_\mu\pa_\nu\phi\right)
\eea
 and the corresponding auxiliary field
 \bea
 F_{\text{G,4}}&=&\frac{2}{(\square\phi)^2-\pa^\mu\pa^\nu\phi\pa_\mu\pa_\nu\phi}\left(\pa^\mu\psi^\alpha\pa_\mu\psi_\alpha\square\phi-\pa^\nu\psi^\alpha\pa^\mu\psi_\alpha\pa_\mu\pa_\nu\phi\right)\label{auxgal4}.
 \eea

 
  In the light of these results we can enlarge our previous families of Lagrangians with extra terms without generating dynamical auxiliary fields (type III). In this case, we allow for the existence of terms of the form $\pa F$ but, as we have seen above, if they appear linearly, they can be eliminated up to a total derivative
 \bea
 \mathcal{L}^{\text{type III}}&=&\int d^2\theta\left( D_\alpha\Phi \mathcal{O}_\beta\Phi\left(\alpha P^{\alpha\beta}(\Phi,D^2\Phi, D^\gamma D^\delta\Phi,\pa^{\gamma\delta}\Phi)+\beta Q^{\alpha\beta}(\pa,\Phi)\right)+\gamma W(\Phi)\right)+\nonumber\\
 &+&\int d^2\theta M\left(\Phi,D^2\Phi, D^\gamma D^\delta\Phi,\pa^{\gamma\delta}\Phi\right),\quad[ \mathcal{O}_\beta]\in\{1,3\} \label{generalaction}
 \eea
 where $Q^{\alpha\beta}$ is an analytical function of the superfield and space-time derivatives, but not of single $D^2$-operators.  The function $P^{\alpha\beta}$ can only generate algebraic expressions of $F$, while $Q^{\alpha\beta}$ can only generate linear terms containing space-time derivatives of $F$. An observation is in order: If the e.o.m. for $F$ cannot be solved for $F$ (for example if $F$ appears linearly in the action) then it can become dynamical (see the example in section V.A). Therefore, we assume that in the expression above, after all possible integration by parts the action in a nonlinear function of $F$.

 
 \subsection{Infinitely many extensions}
 
 In the previous section, the super Galileon action has been obtained in the bosonic sector once we used the trivial solution for $F$, i.e. $F=0$. These solutions are the simplest extensions in the sense that the fermionic sector contains a minimal number of terms. Since the supersymmetric extension is not unique we can look for other extensions such that the Galileon model is obtained on-shell for a non trivial solution of $F$. The result lies in the following fact 
  \begin{lemma}
\label{infextensions}
All three dimensional single scalar field models possess infinitely many $N=1$ supersymmetric extensions.
\end{lemma}
 
  The proof of this lemma is constructive and provides automatically the supersymmetric extension. Let us start with the following supersymmetric type III Lagrangian
 \be
 \mathcal{L}=\int d^2\theta\left( D^\alpha\Phi D_\alpha \Phi\Lambda\left(\Phi,\pa_\mu\Phi,\pa_\mu\pa_\nu\Phi,...\right)+W(\Phi)\right)\label{gensus}
 \ee
 where $\Lambda$ is an undetermined function depending on $\Phi$ and its space-time derivatives and $W$ is a general prepotential. After integration we obtain in the bosonic sector
 \be
 \mathcal{L}\vert_{\psi=0}=\left(-\pa_\mu \phi\pa^\mu\phi+F^2\right)\Lambda(\phi,\pa_\mu,\pa_\mu\pa_\nu\phi,...)+W'(\phi)F\label{genbos}.
 \ee
 
 Solving for $F$
 \be
 F=-\frac{1}{2}\frac{W'(\phi)}{\Lambda(\phi,\pa_\mu,\pa_\mu\pa_\nu\phi,...)}\label{eqF}
 \ee
 and then substituting into (\ref{genbos}), we obtain
 \be
  \mathcal{L}\vert_{\psi=0}=-\pa_\mu \phi\pa^\mu\phi\Lambda(\phi,\pa_\mu,\pa_\mu\pa_\nu\phi,...)-\frac{1}{4}\frac{W'^2(\phi)}{\Lambda(\phi,\pa_\mu,\pa_\mu\pa_\nu\phi,...)}.
 \ee
 
 Now, if $L_g(\phi,\pa_\mu\phi,...)$ is a general Lagrangian (non-supersymmetric) we solve $\Lambda$ for the equation $  \mathcal{L}\vert_{\psi=0}=L_g(\phi,\pa_\mu\phi,...)$ which gives
 \be
 \Lambda(\phi,\pa_\mu,\pa_\mu\pa_\nu\phi,...)_\pm=\frac{-L_g(\phi,\pa_\mu\phi,...)\pm\sqrt{L_g^2(\phi,\pa_\mu\phi,...)-\pa_\mu\phi\pa^\mu\phi W'^2(\phi)}}{2\pa_\mu\phi\pa^\mu\phi}\label{lam}.
 \ee
  
  We replace now $\phi\rightarrow\Phi$ and substitute (\ref{lam}) in (\ref{gensus})
  \be
  \mathcal{L}_\pm=\int d^2\theta\left[ D^\alpha\Phi D_\alpha\Phi\left(\frac{-L_g(\Phi,\pa_\mu\Phi,...)\pm\sqrt{L_g^2(\Phi,\pa_\mu\Phi,...)-\pa_\mu\Phi\pa^\mu\Phi W'^2(\Phi)}}{2\pa_\mu\Phi\pa^\mu\Phi}\right)+W(\phi)\right]\label{tota}.
  \ee
 
 The Lagrangian (\ref{tota}) constitutes a supersymmetric extension of $L_g(\phi,\pa_\mu\phi,...)$ once we solve the e.o.m. for $F$ (\ref{eqF}) with $\Lambda$ given by (\ref{lam}). It is important to note that the prepotential $W(\Phi)$ does not show up in the on-shell bosonic sector (but, of course, contributes to the fermionic sector), therefore any choice of $W$ gives inequivalent supersymmetric extensions with the same bosonic sector (that is why we say infinitely many supersymmetric extensions). In this framework the supersymmetric Galileon models are given by (\ref{tota}) with $L_g(\Phi,\pa_\mu\Phi,...)=\mathcal{L}_{G,i},\phi\rightarrow\Phi$, where $\mathcal{L}_{G,i}$ are the bosonic Galileon Lagrangians (\ref{gal1})-(\ref{gal4}). In Sec. II we have introduced the ``SUSY bosonic twin". The existence of  infinitely many ``SUSY bosonic twins" introduced is Sec. II requires the existence of at least one SUSY version of a given bosonic model (Lemma \ref{infextensions1}). The result contained in Lemma \ref{infextensions} is stronger in the sense that it implies that all bosonic models have a SUSY extension. However there is an important difference between the results provided by Lemma \ref{infextensions1} and Lemma \ref{infextensions}. In the first case, the infinite family of SUSY extensions is related to the canonical SUSY version of the model (this corresponds to $\lambda=0$ in (\ref{susyl1})). In the later, there is no apparent connexion between (\ref{tota}) and its canonical SUSY extension, if any.   
 



\section{The auxiliary field problem}

In the previous sections we have discussed general supersymmetric actions which do not contain (up to total derivatives) terms of the form $\pa F$. Let us assume that the bosonic sector of our action does not contain dynamical terms for the auxiliary field, i.e. F is an algebraic degree of freedom. Now we want to address the following question: can $F$ become dynamical in the fermionic sector? It is easy to construct an action with this property, for example
\be
\mathcal{L}=\int d^2\theta D^\alpha \Phi D_\alpha\Phi D^2\Phi \square \Phi\vert_{\psi=0}=\left( -(\pa\phi)^2+F^2\right)F \square \phi.
\ee

In this case $F$ can be solved algebraically 
\be
F^2=\frac{1}{3}(\pa\phi)^2
\ee
and therefore does not correspond to any physical degree of freedom. However, in the full action, non-trivial derivative terms are generated for the auxiliary field
\bea
\mathcal{L}&=&\left( -(\pa\phi)^2+F^2\right)F \square \phi+\psi^\alpha\psi_\alpha\left((\square\phi)^2+F\square F+i\pa^\alpha_{\,\,\beta}\psi^\beta\square\psi^\alpha\right)\nonumber\\
&+&2\left(i\pa^{\beta\alpha}\phi+C^{\alpha\beta}F\right)\psi_\alpha\left( i\pa^\gamma_{\,\,\beta}\psi_\gamma\square\phi+F\square \psi_\beta\right).
\eea

The presence of these terms leads to a non-algebraic equation for $F$. We can write this equation as follows
\be
F=f(\omega \,\square F , \eta).
\ee

 Since we are considering actions without derivatives of $F$ in the bosonic sector, the parameter $\omega$ is fermionic. In order to see why $F$ remains non-dynamical let us start considering a simple example. We assume that the field equation for $F$ can be written in the from
\be
F=\pa_\mu F\omega^\mu+\eta\label{eqfun},
\ee
where $\omega^\mu$ is a pure fermionic parameter. We have the following sequence:
\bea
\pa_\mu F&=& \pa_{\mu\nu} F\omega^\nu+\pa_\nu F\pa_\mu \omega^\nu+\pa_\mu\eta\label{rec1}\\
\pa_{\rho\mu} F&=&\pa_{\rho\mu\nu}F\omega^\nu+\pa_{\mu\nu}F\pa_\rho\omega^\nu+\pa_{\rho\nu}F\pa_\mu\omega^\nu+\pa_\nu F\pa_{\rho\mu}\omega^\nu+\pa_{\rho\mu}\eta\label{rec2}\\
\pa_{\sigma\rho\mu} F&=&\pa_{\sigma\rho\mu\nu}F\omega^\nu+\pa_{\rho\mu\nu}F\pa_\sigma\omega^\nu+\pa_{\sigma\mu\nu}F\pa_\rho\omega^\nu+\pa_{\mu\nu}F \pa_{\sigma\rho}\omega^\nu+\nonumber\\
&+&\pa_{\sigma\rho\nu}F\pa_\mu\omega^\nu+\pa_{\rho\nu}F\pa_{\sigma\mu}\omega^\nu+\pa_{\sigma\nu}F\pa_{\rho\nu}\omega^\nu+\pa_\nu F \pa_{\sigma\rho\mu}\omega^\nu+\pa_{\sigma\rho\mu}\eta\label{rec3}
\eea
where $\pa_{\mu\nu\rho...}\equiv\pa_\mu\pa_\nu\pa_\rho...$.  Due to the fermionic nature of $\omega^\mu$ we have the following identity in three dimensions 
\be
\Omega^{\mu\nu\rho\sigma}\omega_\mu\omega_\nu\omega_\rho\omega_\sigma=0,
\ee
i.e. all combinations with one repeated index give zero. We assume from now that $\omega_i^2=0$. In the presence of more Grassmann coordinates, for example in a model with more than one superfield, the nilpotent index for $\omega_\mu$ can be higher, but the process for eliminating $F$ will be still finite. In three dimensions we can have $\omega_0\omega_1\omega_2$, but the next term added to the product necessarily repeats one index. As a consequence the higher derivative term in (\ref{rec3}) does not appear in (\ref{eqfun}) if we substitute successively (\ref{rec1})-(\ref{rec3}). So we need only up to third order terms to solve the equation (fourth order terms are supressed by the combination $\omega_\mu\omega_\nu\omega_\rho\omega_\sigma$ ).  If we continue the substitutions the third order terms appear accompanied by $\pa_\mu\omega_\nu$. We have the identity
\be
\Omega^{\mu_1\nu_1...\mu_{10}\nu_{10}}\pa_{\mu_1}\omega_{\nu_1}...\pa_{\mu_{10}}\omega_{\nu_{10}}=0.
\ee

Therefore after at most ten substitutions the third order terms disappear. The second order terms are accompanied by $\pa_{\mu}\omega_\nu$. The number of different terms of the form $\pa_{\nu_1}...\pa_{\mu_n}\omega_\nu$ is given by
\be
\#_{d,n}\equiv\# \left(\text{different terms}\,\, \pa_{\nu_1}...\pa_{\mu_n}\omega_\nu\right)=\left(\begin{matrix}
    d+n-1   \\
     n
\end{matrix}\right)\,d
\ee
where $d$ is the dimension of the space. This can be straightforwardly generalized if in the original equation enters a tensor of rank $N$

\be
\#_{d,n}^N\equiv\# \left(\text{different terms}\,\, \pa_{\nu_1}...\pa_{\mu_n}\omega_{\nu_1,...\nu_N}\right)=\left(\begin{matrix}
    d+n-1   \\
     n
\end{matrix}\right)\,d^N
\ee

 Since $\#_{3,2}=18$ in at most 19 substitutions all second order terms disappear. As a conseqence no more higher derivative terms are generated this process ends always in a finite number of steps, remaining F non-dynamical. Let us consider now the most general situation. Under our assumption there are no terms with derivatives in $F$ in the bosonic sector. We have
\be
f(F)=h(\mathcal{O}_1F\omega_1, \mathcal{O}_2F\omega_2, ..., \eta)
\ee 
where $\mathcal{O}_i$ are differential operators and $\omega_i$ fermionic parameters. We can invert formally the equation
\be
F=f^{-1}h(\mathcal{O}_1F\omega_1, \mathcal{O}_2F\omega_2, ..., \eta).
\ee

We assume that $\mathcal{O}_1=\pa_{\mu_1...\mu_n}$ is the highest order differential operator.  Now we determine the sequence
\be
\pa_{\mu_1} F,...,\pa_{\mu_1...\mu_n} F\equiv \mathcal{O}_1 F, \pa_{\nu_1} \mathcal{O}_1 F,\pa_{\nu_1}\pa_{\nu_2} \mathcal{O}_1 F,...\label{finit}.
\ee

The highest derivative $\max (\mathcal{S})$ of $F$ we have to reach in the sequence will be
\be
\max (\mathcal{S})=\sum_{i=1}^n \,i\,\#_{d,n-i}^N=\sum_{i=1}^n\,i\,    \left(\begin{matrix}
    d+n-1-i   \\
     n-i
\end{matrix}\right)d^N\label{genfor},
\ee
once we reach this order the equation is ``closed" in the sense that no more higher derivative terms are generated. After that, we have to substitute a finite number of times the terms in the sequence (\ref{finit}) until all $F$ derivatives are eliminated. We can state this result in the following lemma

\begin{lemma}
\label{auxfieldproblem}
If the e.o.m. for the auxiliary field $F$ is algebraic in the bosonic sector, it remains algebraic in the full supersymmetric action.
\end{lemma}

We have shown that the equations for $F$ are solved in a finite number of steps, but even for the simplest cases, the explicit form of the solution is not very enlightening (see Appendix B). This result can be straightforwardly generalized to four dimensions. In the light of these results we can write down the most general $N=1$ without dynamical auxiliary fields
 \bea
 \mathcal{L}^{\text{type IV}}&=&\int d^2\theta D_\alpha\Phi \mathcal{O}_\beta\Phi\left(\alpha P^{\alpha\beta}(\Phi,D^2\Phi, D^\gamma D^\delta\Phi,f(\pa)\Phi)+\gamma W(\Phi)\right)+\nonumber\\
 &+&\int d^2\theta M\left(\Phi,D^2\Phi, D^\gamma D^\delta\Phi,\pa^{\gamma\delta}\Phi\right),\quad[ \mathcal{O}_\beta]\in\{1,3\} \label{generalactionm}.
 \eea
 
 We are assuming  again that if we consider (\ref{generalactionm}) as a polynomial in $F$, it is at least quadratic.


\subsection{Propagating auxiliary fields in the bosonic sector}

It is simple to construct supersymmetric actions containing derivatives of the auxiliary field in the bosonic sector. For example, if a spacetime derivative shows up in the superfield action we can ensure that $F$ will appear with derivatives. If this happens, the auxiliary field can become dynamical, and since fermionic and bosonic degrees of freedom must be balanced in a supersymmetric theory,  both components of the spinor would now propagate generating a real scalar superfield with two bosonic and two fermionic d.o.f. If one of the extra bosonic d.o.f. is ghost-like then it must be paired to a ghost-like fermionic state.

  Let us consider for example the following model
\be
\mathcal{L}=\frac{1}{2}\int d^2\theta \pa_\mu\Phi\pa^\mu \Phi=-\pa_\mu \phi\pa^\mu F+\frac{1}{2}\pa_\mu\psi^\alpha\pa^\mu\psi_\alpha\label{gh1}.
\ee

A simple Hamiltonian analysis leads to four degrees of freedom (two bosonic and two fermionic). After the replacement
\be
\phi\rightarrow A+B,\quad F\rightarrow A-B, 
\ee
we can rewrite the action as follows
\be
\mathcal{L}=-\pa_\mu A\pa^\mu A+\pa_\mu B\pa^\mu B+\frac{1}{2}\pa_\mu\psi^\alpha\pa^\mu\psi_\alpha.\label{gh}
\ee

In (\ref{gh}) we identify clearly a ghost degree of freedom ($B$), and therefore the presence of derivatives in F, in this case, leads to ghost terms. After integrating by parts in (\ref{gh1}), we see that $F$ plays the role of a Lagrange multiplier imposing the field equation for $\phi$ ($\square\phi=0$). The generalization of this fact is straightforward. Let us take a general single field bosonic action
\be
\mathcal{L}(\phi,\pa_\mu\phi,\pa_{\mu\nu}\phi,...).\label{gen1}.
\ee 

If the equation of motion for $\phi$ in (\ref{gen1}) is $\mathcal{E}(\phi,\pa_\mu\phi,\pa_{\mu\nu}\phi,...)=0$ we have
\be
\int d^2\theta \mathcal{L}(\Phi,\pa_\mu\Phi,\pa_{\mu\nu}\Phi,...)=F\,\mathcal{E}(\phi,\pa_\mu\phi,\pa_{\mu\nu}\phi,...)+\text{fermions}
\ee
and therefore the variation with respect to $F$ gives the equation of motion of the field $\phi$ in the original theory \cite{q7}. The variation with respect to $\phi$ gives a differential equation for $F$. In general, for this kind of models, each supermultiplet contains 4 instead of two degrees of freedom,  being two of them (one bosonic and one fermionic) ghost-like.

\section{Summary}

In this paper we have studied supersymmetric higher-derivative field models in three dimensions. First, we have shown how to construct all possible $N=1$ models with non-trivial bosonic sector and classified them in two families. By restricting the degree of the operators acting on the supefields we have provided a classification of all models without derivatives acting on the auxiliary field (up to a total derivative). In particular, we have proposed a SUSY formulation of $P(X)$-theories where $F$ can be eliminated algebraically even in the full theory. 

In three dimensions the most general Galileon theory is a linear combination of three terms, namely: the quadratic Galileon (the linear $\sigma$-model term), the cubic and the quartic Galileons (the proper higher derivative term). The SUSY version of these theories turns out to be very simple (in contrast to what happens in four dimensions \cite{Khoury2, Koehn2, Farakos}) due to the key structure of the SUSY linear $\sigma$-model term $D^\alpha \Phi D_\alpha\Phi$. This term ``saturates" the Grassmann integration implying that any function of the superfields multiplying this term will show up in the bosonic sector in its lowest component ($\theta\rightarrow0$). Since all Galileon theories are of the form $(\pa\phi)^2 f(\phi,\pa_\mu\phi,...)$, the SUSY extension is almost straightforward. It is important to note that in four dimensions (or in three dimensions and extended supersymmetry) the trivial K\"{a}hler potential ($K(\Phi^\dagger\Phi)=\Phi^\dagger\Phi$) leading to the linear $\sigma$-model term does not have this property. Our SUSY extension is ghost-free and satisfies the (obvious) supersymmetric generalization of the Galilean shift. 

It is well-known that the SUSY extension of a given bosonic theory is not unique. We have shown that, for scalar field models in three dimensions and $N=1$, SUSY versions are not only non-unique, but there are infinitely many possible extensions. Since the proof is constructive, we also have shown how to construct these supersymmetric actions. In particular we were able to build an infinite number of non-equivalent SUSY extensions of the Galileon theories. These extensions do not respect the super-Galileon shift, but they are still free of ghosts. 

In all supersymmetric models the auxiliary field plays an important role. In some SUSY models its trivial solution provides the correct bosonic sector but, as we have shown in section IV.A sometimes the non-trivial solution for $F$ gives the correct bosonic sector modifying the kinetic terms (note that usually, for example in nonlinear $\sigma$-models with potentials, $F$ does not change the kinetic Lagrangian). One way or another, $F$ can be eliminated algebraically. In the last section we have discussed what happens when $F$ appears with derivatives in the fermionic sector. At first glance, it might seem that $F$ becomes dynamical, but we have shown that due to the Grassmann nature of the fermionic sector it can be eliminated algebraically.  The situation is different when $\pa F$-terms show up in the bosonic sector. Of course, $F$ can be dynamical (with the appropriate terms). Sometimes this extra degree of freedom is well-behaved \cite{Nitta7}, but it can be also a ghost state, as we have seen in several examples.

The role of the auxiliary field in SUSY theories (especially when it becomes dynamical) and its relation to ghost degrees of freedom is under current investigation.

{\bf Acknowledgements.}- The author thanks Prof. A. Wereszczynski for useful discussions and important improvements in a previous version of the manuscript. 

\appendix

\section{Useful identities}

From the form of the superderivative (\ref{superd}) we have the following identities
\bea
D_\alpha D_\beta&=&i\pa_{\alpha\beta}+C_{\beta\alpha}D^2\\
D_\alpha D_\beta D^\alpha&=&0\\
D_\gamma D_\alpha D_\beta&=&i\pa_{\alpha\beta}D_\gamma-i C_{\beta_\alpha}\pa_{\gamma\delta}D^\delta\\
D^2 D_\alpha&=&i\pa_{\alpha\beta}D^\beta\\
D^2 D^2&=&\square	\label{idap5}\\
D_\alpha D_\beta D_\gamma D_\delta&=&-\pa_{\alpha\beta}\pa_{\gamma\delta}+i\left(\pa_{\alpha\beta}C_{\delta\gamma}+\pa_{\gamma\delta}C_{\beta\alpha}\right)D^2	+C_{\beta\alpha}C_{\delta\gamma}\square.
\eea

The operators with odd degree are proportional to the $D$-operator, and since $D^\alpha \Phi\vert=\psi^\alpha$, they only generate fermions. On the other hand, if the degree of the operator is even, it can be written as a combination of spacetime derivatives $\pa$ and $D^2$-operators, and since $\pa \Phi\vert =\pa\phi$ and $D^2 \Phi\vert=F$ these terms only generate bosonic terms. In Table \ref{tab1} we summarize the algebra of operators. The symbol  $\langle1,D^2\rangle$ stands for the space generated by the even operators $\pa$ and $D^2$.
The spinor superfield ($D_\alpha \Phi$) can be expanded as follows
\be
D_\alpha \Phi=\psi_\alpha-\theta_\alpha F+i\theta^\beta\sigma^\mu_{\alpha\beta}\pa_\mu\phi+i\theta^\beta\theta^\gamma\sigma^\mu_{\beta\alpha}\pa_\mu\psi_\gamma.
\ee

With our conventions we have
\bea
\int d^2\theta \theta^2&=&-1\\
\theta^2&\equiv&\frac{1}{2}\theta^\alpha\theta_\alpha\\
\theta^\alpha\theta^\beta&=&C^{\alpha\beta}\theta^2\\
\text{tr}\,\sigma^\mu\sigma^\nu&=&2\eta^{\mu\nu}.
\eea

\begin{table}
   \caption {D-algebra} \label{tab1} 
\begin{center}
  \begin{tabular}{ l | l | l | l l | l}
    & $\pa$ & $D $&$ D^2$ \\ \hline
    $\pa$ & $\pa$ & $D$& $D^2$ \\ \hline
    $D$ & $D $& $\langle\pa,D^2\rangle$ &$ D$\\
    \hline
    $ D^2 $&$ D^2$ & $D$ & $\square$ \\
  \end{tabular}
\end{center}
\end{table}

\section{Explicit solutions for $F$: simple examples}

As we have seen in Sect. V, the number of steps to close the equation for $F$ grows with the dimension of the space and with the number of derivatives acting on $F$. The simplest example we can study is the following unidimensional problem
\be
F=F' \omega +\eta\label{sim}
\ee

From (\ref{genfor}) we have that the maximum order of the derivative in the sequence (\ref{finit}) will be $\max(\mathcal{S})=1$ ($d=1, n=1, N=1$). By differentiating in (\ref{sim}) we obtain
\be
F'=F''\omega+F'\omega'+\eta'.
\ee

Now, since after substituting in (\ref{sim}) the term proportional to $F''$ vanishes, the equation is closed. If we repeat the process twice we arrive at the solution 
\be
F=\eta\omega'\omega+\eta' \omega+\eta.
\ee

Let us assume now that the field equation for $F$ takes the simple form
\be
F=\pa_\mu F\omega^\mu+\eta.
\ee

 If we apply (\ref{genfor}) we get $\max(\mathcal{S})=2$ in two dimensions ($d=2, n=1, N=2$). This can be immediately confirmed by means of (\ref{rec1})-(\ref{rec2}). But even if the equation is closed at order 2, the explicit form for $F$ is extremely large
 \bea
 F&=&\eta+\omega^\mu\pa_\mu\eta+\omega^\mu\omega^\nu\pa_{\mu\nu}\eta+\omega^\mu\pa_\mu\omega^\nu\pa_\nu\eta+\omega^\mu\pa_\mu\omega^\nu\pa_\nu\omega^\rho\pa_\rho\eta\nonumber\\
 &+&\omega^\mu\omega^\rho\pa_\mu\omega^\nu\pa_\nu\omega^\epsilon\pa_{\rho\epsilon}\eta+...(\text{156 terms})+...\omega^\mu\omega^\epsilon\pa_\rho\omega^\tau\pa_\mu\omega^\nu\pa_\mu\omega^\rho\pa_\epsilon\omega^\alpha\pa_{\tau\alpha}\omega^\beta\pa_\beta\eta.
 \eea
 
 In order to illustrate how $\max(\mathcal{S})$ grows let us take the following equation in four dimensions

 \be
 F=\pa_\mu\square F \omega^\mu+\eta
 \ee
 
 In this case, the sequence (\ref{finit}) closes at order $\max(\mathcal{S})=84$ and as we have seen in the previous examples we have still to substitute derivatives up to 84 order until all terms $\pa...\pa F$ are eliminated.

\end{document}